\documentclass[preprint,12pt]{elsarticle}

\journal{Annals of Physics}

\usepackage{graphicx}
\usepackage[english]{babel}
\usepackage{bm}
\usepackage{amsmath,amsfonts}
\usepackage{amsbsy}
\usepackage[colorlinks=true,linkcolor=blue,urlcolor=blue,citecolor=blue,breaklinks=true]{hyperref}
\usepackage[utf8]{inputenc}

\usepackage{graphics}
\usepackage{subcaption}
\usepackage{xcolor}
\usepackage{bbm}
\usepackage[normalem]{ulem}
\usepackage{amssymb}
\usepackage{array}
\usepackage{amsmath}
\usepackage{graphicx}\usepackage{graphics}
\usepackage{dcolumn}
\usepackage{bm}\usepackage{varioref}
\begin{document}

\begin{frontmatter}
\title{Current-voltage  characteristics of superconductor-normal metal-superconductor junctions}

\author{T.  Liu}
\author{A. V.  Andreev}
\author{B. Z.  Spivak}

\address{Department of Physics, University of Washington, Seattle, WA 98195,  USA}

\date{\today}

\begin{abstract}
 {\it  Dedicated to  the memory of Kostya Efetov, a great physicist and friend.  }
\smallskip \\
We develop a theory of current-voltage (I-U) characteristics for superconductor-normal metal-superconductor (SNS) junctions.
At small voltages and sufficiently low temperatures  the I-U characteristics of the junction is controlled by the inelastic relaxation time $\tau_{in}$. In particular, the linear conductance  is proportional to $\tau_{in}$.  In this regime  the I-U characteristics can be expressed solely in terms of dependence of the density of states in the normal region  $\nu(\chi)$ on the phase difference of the order parameter across the the junction.
In contrast,  at large voltages the I-U characteristics of the device is controlled by the elastic relaxation time $\tau_{el}$, which is much smaller than the inelastic one.
\end{abstract}

\end{frontmatter}

\section{Introduction}

\begin{figure}[!ht]
\captionsetup[subfigure]{labelformat=empty}
\subfloat[]{\label{SNSFIG_A}\includegraphics[scale = 0.3]{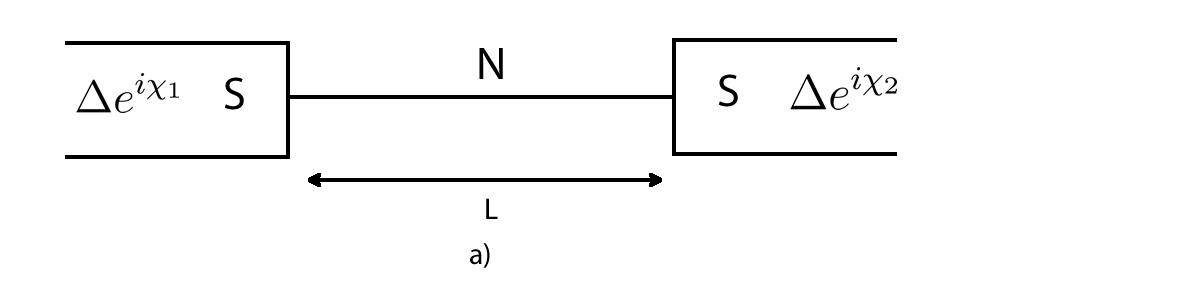}} \\
\subfloat[]{\label{SNSFIG_B}\includegraphics[scale = 0.3]{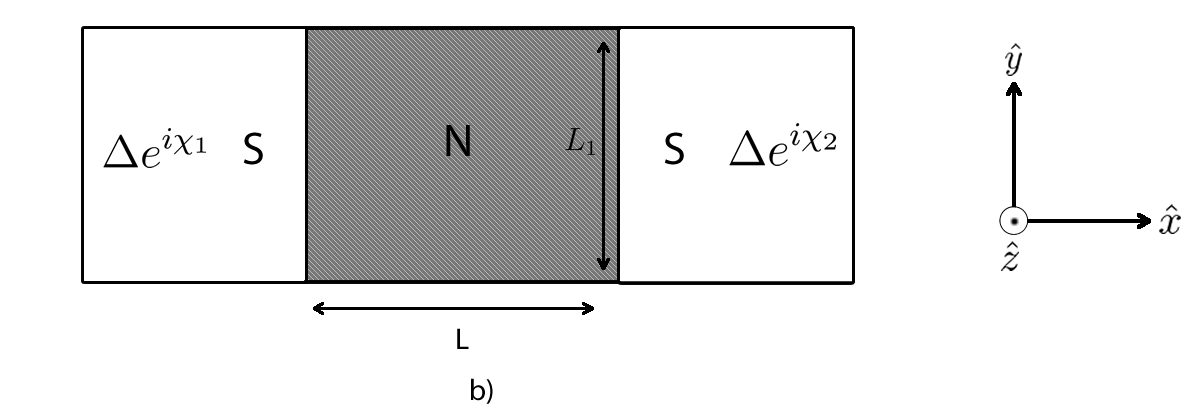}} \\
\subfloat[]{\label{SNSFIG_C}\includegraphics[scale = 0.3]{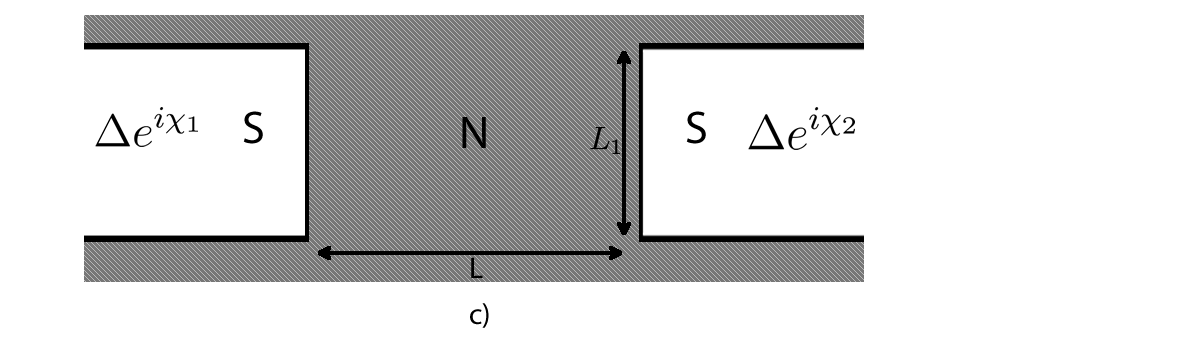}}
\caption{Qualitative representation of  a) 1D SNS junction b) Bulk junction with closed boundaries c) Bulk junction with open boundaries.}
\label{SNSFIG}
\end{figure}

The theory of current-voltage (I-U) characteristics of  superconducting weak links  at relatively large voltages has been developed in many articles (see for example \cite{LarkinOvchinnikov1,Tinkham,Volk,Averin1},  and references therein). However at small voltages  the I-U characteristics exhibit interesting features which are quite different from those at large voltages, this regime attracted much less attention. In this article we focus on the theory of I-U characteristics of SNS junctions in this regime.   
A schematic picture of an SNS junction in which the normal metal section of the junction is 
 sandwiched in between two s-wave superconductors,
 is presented in Fig.~\ref{SNSFIG}.

The difference between the phases of the order parameter on different sides on the junction $\chi=\chi_{1}-\chi_{2}$ is related to the voltage across  the junction $U$ by the Josephson relation,
\begin{equation}\label{eq:Joshepson}
\frac{d\chi}{dt}  = 2eU(t).
\end{equation}

The most general description of quantum systems is in terms of the statistical matrix (or many-body density matrix) $\hat{w}$.  Let us represent this matrix in the basis of eigenstates for the  instantaneous Hamiltonian $\hat{H}  (t) $. The expectation value of the current operator, $\langle J\rangle = \mathrm{tr} (\hat{J} \hat{w})$, may be written as
\begin{equation}\label{eq:J}
\langle J\rangle = \sum_n w_{nn}J_{nn} + \sum_{n\neq m} w_{nm}J_{mn}= J_{d}+J_{nd}.
\end{equation}

 Here the first term represents the diagonal contribution to the current, and the second term represents the non-diagonal  contribution. In particular, in thermal equilibrium, where the statistical matrix is given by the Gibbs distribution, $\hat{w}= \exp (-\beta \hat{H})/Z$, with $Z$ being the partition function, the diagonal contribution corresponds to the equilibrium current. 
A canonical example of the diagonal component, $J_{d}$, is the equilibrium  super-current  in superconductors. We note that in non-equilibrium situations $J_{d}$ contains both the dissipative and non-dissipative parts.
In a situation where the statistical matrix contains non-diagonal elements, the expectation value of the current acquires a non-diagonal contribution,  $J_{nd}$.
An example of the non-diagonal component, $J_{nd}$, is the ohmic current in normal metals. In this case, according to the Kubo formula,  $J_{nd}$ is related to transitions between electronic eigenstates induced by the external electric field.

%
We show below that at small voltages in an SNS junctions, $J_{d}\gg J_{nd}$, the diagonal component of the current controls both the dissipative and non-dissipative part of the current. The reason for this is that the dissipative part of $J_{d}$ is proportional to the inelastic mean free time $\tau_{in}$, while $J_{nd}$ is proportional to the elastic one $\tau_{el}$ , which is usually  much shorter than $\tau_{in}$. In this regime, $J_{d}$ can
be evaluated in the adiabatic approximation, and it can be expressed in terms of the phase $\chi$ and energy $\epsilon$ dependence of the quasi-particle density of states in the normal part of the junction $\nu(\epsilon,\chi)$.

The physical origin of this contribution to the current is similar to the Debye mechanism of microwave absorption in gases \cite{Debye}, Mandelstam-Leontovich mechanism of the second viscosity in liquids \cite{Landau}, the Pollak-Geballe mechanism of microwave absorption in the hopping conductivity regime \cite{Pollak}, and the mechanism of low frequency microwave absorption in superconductors \cite{Smith1, Smith2}.

In principle, such a mechanism exists independently of the nature of electronic states in the normal region of SNS junctions. It is also valid in the case where the electronic state in the normal region is strongly correlated; for example, the quantum Hall states \cite{Yacobi,Finkelstein}. In this article, however, we restrict ourselves to the case where the exited states of the electronic liquid can be described by system of Fermionic quasi-particles.  
 
{ The I-U characteristics of SNS junctions depend on the external circuits to which they are connected.  In what follows, we will be interested in I-U characteristics 
of the junctions in situations where either the voltage (voltage bias setup) or current (current bias setup) is fixed by the external circuit. 
In Figs.~\ref{FigfixedV_IVC.pdf},~\ref{Fig_Ut.pdf}, and~\ref{IVC_fixedI.png} we qualitatively summarize our results for the cases of voltage-  and current-biased junctions. 

In the case of voltage-biased  junctions the I-U characteristic turns out to be non-monotonic, and the maximum current $J_{max}$ is reached at $eU\sim \tau_{in}^{-1}$ \cite{AslLark}. We will show that the value of $J_{max}$ can be significantly larger than the temperature-dependent critical current $J_{c}(T)$, and in some cases it can be as large as the zero temperature critical current  $J_{c}(0)$. At even larger voltages the I-U characteristic reaches a minimum, after which the current increases with voltage.

In the case of  current-biased  junctions at $J_{c}<J<J_{jump}$ the voltage monotonically increases from zero to a relatively small value, which is inversely proportional to $\tau_{in}$. Then, at  $J=J_{jump}\sim J_{max}$ the I-U characteristic exhibits a  jump to a significantly higher voltage.

The presentation below is organized as follows. In Sec.~\ref{sec:level-dynamics} we 
obtain general expressions for the diagonal contribution to current in terms of the inelastic relaxation time $\tau_{in}$ and sensitivity of quasi-particle energy levels to the change in the phase difference across the junction. In Sec.~\ref{sec:IU_general} we discuss the characteristic features of the current-voltage characteristics of voltage and current-biased SNS junctions, which are caused by the presence of the long inelastic relaxation time, $\tau_{in}$ in the system. In Sec.~\ref{sec:Character} we apply the general formalism developed in Sec.~\ref{sec:level-dynamics} to study the I-U characteristics of ballistic single channel junctions (Sec.~\ref{sec:clean}) and diffusive multi-channel junctions (Sec.~\ref{diffusive case}).  We present our conclusions in Sec.~\ref{sec:conclusions}.
Finally, in \ref{sec:derivationLO} we present a derivation of our general equations in Sec.~\ref{sec:level-dynamics} in the diffusive regime starting from the Larkin-Ovchinnikov equations for the quasi-classical Green's functions.

\section{Description of the dynamics of SNS junction in adiabatic approximation.
\label{sec:level-dynamics}}

Due to Andreev reflection from the normal metal-superconductor boundaries of the  SNS junction, low energy ($\epsilon< \Delta$) quasi-particles  are trapped inside the normal region.
If the voltage across the SNS junction is sufficiently small, the quasi-particle energies $\epsilon_{i}(\chi(t))$
can be calculated in the adiabatic approximation, treating the phase difference $\chi(t)$ as a parameter.

At finite temperature, the quasi-particles occupying these levels move in energy space together with the levels. This motion creates a non-equilibrium quasi-particle distribution, which relaxes via inelastic scattering and leads to dissipation. There are two equivalent ways to describe this non-equilibrium distribution. The first is to describe the occupancy of time-dependent energy levels. This description is similar to the Lagrangian description of fluid dynamics.  The second approach is to consider the electron distribution as a function of energy, in analogy to the Eulerian description of fluid dynamics.

The Lagrangian description is convenient in the cases where individual quasi-particle  energy levels are well resolved, and the Eulerian description is more suitable for systems where energy levels form a continuum. In order to obtain the kinetic description of non-equilibrium dynamics of the junctions it is easier to start with the Lagrangian description. The corresponding equations in the Eulerian approach are  then obtained by a straightforward change of variables.

\subsection{Lagrangian description of dynamics of SNS junctions. }

 Let us introduce the occupation number of   $i_{th}$ level $n_{i}(t)$. In the adiabatic approximation only scattering can change the occupation of a particular level, so the time evolution of $n_i (t)$ is controlled by the following equation,
\begin{equation}\label{eq:LagrKinEq}
\frac{d n_{i}(t)}{d t}=I_{st} \{n_{i}\}.
\end{equation}
We will use an expression for the scattering integral in the relaxation time approximation
\begin{equation}\label{eq:LagrKinEqST}
I_{st} = \frac{ n_F(\epsilon_i(t)) -  n_i(t)}{\tau_{in}},
\end{equation}
where $n_{F}(\epsilon)=1/(1+\exp(\epsilon/T))$ is the Fermi distribution function, and we assume that the relaxation time  $\tau_{in}(T)$ depends only on the  temperature.

In general, the relaxation time approximation is valid with precision of order one.
However, in some cases this approximation turns out to be asymptotically exact. In particular, this is the case when the normal part of the junction is in the diffusive limit and the temperature is  larger than the Thouless energy. (See the corresponding discussion in Section \eqref{diffusive case})
At $t\gg \tau_{in}$ the general solution of Eqs.~\eqref{eq:LagrKinEq},\eqref{eq:LagrKinEqST} is given by

\begin{equation}\label{nGenL}
n_i(t) = \int^\infty _ 0 \frac{d\tau}{\tau_{in}}  e^\frac{-\tau}{\tau_i} n_F (\epsilon_i(t-\tau)).
\end{equation}
The diagonal component of the current through the junction  can be written as
\begin{equation}\label{eq:J_Lagr}
 J_{d}  = 2e\frac{\partial E}{\partial \chi}
=  J_c(0) Y(\chi,0)+ 2e\sum_{i}
\frac{\partial \epsilon_{i}(\chi)}{\partial \chi}n_{i}.
 \end{equation}
The first term in Eq.~\eqref{eq:J_Lagr} represents the super-current through the system in the ground state. Here $J_{c}(0)$ is the critical current at zero temperature, and $Y(\chi, 0)$ is a periodic function with maximum $1$ and a period $2\pi$.

\subsection{Eulerian description}

In the Eulerian description  the quasi-particle  distribution function inside the normal region is a function of energy and time $n(\epsilon, t)$.
This description is convenient in the case where the energy levels are broadened on the energy scale larger than the level spacing. The number of levels in the system is conserved, so the density of states is therefore subject to the continuity equation in energy space
\begin{equation}\label{eq:levelContinuity}
\partial_t \nu(\epsilon, \chi) + \partial_\epsilon \big ( v_{\nu} (\epsilon, \chi)\nu (\epsilon, \chi) \big)  =0,
\end{equation}
where $v_\nu (\epsilon,\chi)$ is the level ``velocity'' in energy space. Using Eqs.~\eqref{eq:Joshepson},~\eqref{eq:levelContinuity} the level velocity can be expressed in the form
\begin{equation}\label{eq:level_continuity}
v_{\nu} (\epsilon, \chi)= 2 e U \cdot V_{\nu}(\epsilon, \chi),
\end{equation}
 where
\begin{equation}\label{eq:level_velocity}
V_{\nu} (\epsilon,\chi) = - \frac{1}{\nu (\epsilon, \chi)}  \int_{0}^{\epsilon} d \tilde{\epsilon}  \frac{\partial \nu (\tilde{\epsilon}, \chi)}{\partial \chi}
\end{equation}
characterizes the sensitivity of the energy levels to changes of  $\chi(t)$.
In the absence of inelastic scattering, the time evolution due to the spectral flow is described by the continuity equation $\partial_t(\nu n)+\partial_\epsilon(v_{\nu} \nu n) =0$. Combining it with Eq.~\eqref{eq:level_continuity} for $\nu (\epsilon, \chi)$ and allowing for inelastic collisions we obtain the kinetic equation
\begin{equation}\label{EQ:N_DOT}
 \partial_{t} n (\epsilon, t)+ 2 e U(t)\cdot  V_{\nu} (\epsilon, \chi)\,  \partial_\epsilon  n(\epsilon, t) = I_{\mathrm{in}}\{  n\} .
\end{equation}
The expression for the current in the Eulerian description has a form
\begin{equation}\label{EQ:CURRENTEUL}
J_{d}=J_{c}(0)Y(\chi, 0) -2e \int_{0}^{\infty} d \epsilon \nu(\epsilon, t) n(\epsilon, t) V_{\nu}(\epsilon, \chi).
\end{equation}
Introducing the integrated density of states
\begin{equation}
N(\epsilon,t) = \int^\epsilon_0 d \epsilon \nu(\epsilon,t),
\end{equation}
and changing the variables from $\epsilon$ to $N(\epsilon,t)$, we can write Eq.~\eqref{EQ:N_DOT} as
\begin{equation}
\partial_t n(N,t) = \frac{n_F(\epsilon(N,t)) - n(N,t)}{\tau_{in}}.
\end{equation}
 It has a general solution given by,
\begin{equation}\label{nGenE}
n(N,t) = \int^\infty _ 0 \frac{d\tau}{\tau_{in}}  e^\frac{-\tau}{\tau_{in}} n_F (\epsilon(N,t-\tau)).
\end{equation}

\emph{Small voltage regime:}
The description of dissipative current presented above simplifies significantly for slow time-dependence of the phase difference,  $\dot{\chi}(t) =2eU (t)  \ll \tau_{in}^{-1}$.  In this case, to first order accuracy in $U(t)$, the diagonal contribution to the current can be written in the form
\begin{equation}\label{eq:G_chi_def}
 J_{d} (t, T)=J_{c}(T)Y(\chi(t),  T)+G_{d}[\chi(t)] U(t).
\end{equation}
Here the first term represents  the equilibrium super-current corresponding to the instantaneous value of $\chi(t)$.  It is convenient to express it as a product of the temperature dependent critical current $J_{c}(T)$ and a dimensionless periodic function periodic function of $\chi$ of unit amplitude,  $Y(\chi, T)$.  For example,  at large temperatures $Y(\chi, T)\sim \sin \chi$.

The second term in Eq.~\eqref{eq:G_chi_def} describes the diagonal contribution of the dissipative current and  is characterized by the ``diagonal conductance'' $G_{d}[\chi(t)]$,  which depends on the instantaneous phase difference  phase difference $\chi(t)$. It can be evaluated by solving  Eqs.~\eqref{eq:LagrKinEq}, ~\eqref{eq:LagrKinEqST}, and \eqref{EQ:N_DOT}  to first order in $U(t)$, then substituting the result into equation Eqs.~\eqref{eq:J_Lagr},~\eqref{EQ:CURRENTEUL}. 
This yields the following expressions for the diagonal conductance in the Lagrangian and Eulerian variables
\begin{subequations}\label{eq:condChi_ab}
\begin{eqnarray}\label{eq:condChi_a}
G_{d}[\chi] & = &   -4e^2  \tau_{in} \sum_i \partial_\epsilon n_F(\epsilon_i) \big(\partial_\chi \epsilon_{i}(\chi) \big)^{2}   \\
\label{eq:condChi_b}
 & = & -4e^2 \tau_{in} \int_{0}^{\infty}  d \epsilon \nu(\epsilon, \chi) V^2_{\nu}(\epsilon, \chi)
\partial_\epsilon n_F(\epsilon).
\end{eqnarray}
\end{subequations}
Thus, at sufficiently small voltages the diagonal contribution to the current can be expressed in terms of the phase dependent density of states $\nu(\epsilon,\chi)$, and is proportional to the inelastic relaxation time $\tau_{in}$.

 The non-diagonal contribution to the current corresponds to elastic electron transfer between the superconducting banks of the junction, and  may be expressed as
  $J_{nd}  = G_{nd} U(t) $.  Since  it is not proportional to $\tau_{in}$ we have $G_{nd}\ll G_{d}$.  Therefore,  at small voltages,  it is possible to neglect $J_{nd}$ compared to $J_{d}$.

Equations~\eqref{eq:LagrKinEq}-\eqref{EQ:CURRENTEUL} which describe  slow dynamics of SNS junctions in terms of
  the $\chi$ and $\epsilon$ -dependence of  the quasi-particle density of states $\nu(\epsilon,\chi)$ are quite general.  They hold at relatively small voltages, where  the spectrum of quasi-particles in the normal region of the junction can be calculated  in the adiabatic approximation,
and the quasi-particle distribution function inside the normal region is spatially  uniform.

 In \ref{sec:derivationLO} we present a derivation  of Eqs.~\eqref{eq:level_velocity}, \eqref{EQ:N_DOT}, and \eqref{EQ:CURRENTEUL} in the diffusive regime, $L\gg l$, using a procedure developed by Larkin-Ovchinnikov \cite{LarkinOvchinnikov}. Here $l=v_{F}\tau_{el}$ is the elastic mean free path, $v_{F}$ is the Fermi velocity, and $L$ is the length of the junction (See Fig. \ref{SNSFIG}).

\section{General features of I-U characteristics of SNS junctions}
\label{sec:IU_general}

The form of $\chi(t)$  in a  junction depends on the external circuit.  Below we consider the I-U characteristics for two common setups: voltage-biased junction, and current-biased junction.

We show that the  existence of the  long inelastic relaxation time $\tau_{in}$ has a dramatic effect on the shape  of the I-U characteristics of the junctions.   In the voltage bias case  the I-U characteristic becomes  non-monotonic: it acquires an $N$-shape, as illustrated in Fig.~\ref{FigfixedV_IVC.pdf}.
In the current-bias case the voltage dependence on the applied current is illustrated in Fig.~\ref{IVC_fixedI.png}. Broadly speaking it consists of two regions: 1) At relatively small  excess of the bias current over the critical current  the time-averaged voltage across the junction  monotonically increases from zero, while its value remains rather small (inversely proportional to the inelastic relaxation time),
2) At larger bias currents, $J \sim J_{jump}$, the voltage  exhibits a sharp jump  to a much higher value. This feature of  the dependence of the voltage on the bias current may have important implications for the interpretation of experimental data;  because of the low values of the voltage in region 1)  the transition to region 2) may be mistaken for the transition from the dissipationless to the dissipative state of the junction. Below, we show  that the shape of the I-U characteristics
at low voltages
can be described in terms of the phase-dependence of the quasi-particle density of states in the junction.

\subsection{Voltage biased SNS junctions}
\label{sec:voltage-bias}

In the voltage bias case we define the nonlinear conductance $\bar{G}(U)$ as
\begin{equation}\label{eq:G_constant_U_def}
\bar{G}(U) =   \frac{\langle J (t) \rangle}{U},
\end{equation}
where $\langle \ldots \rangle$ denotes averaging over time.
Since the phase winds at a constant rate via the Josephson relation Eq.~\eqref{eq:Joshepson}, after averaging over time the non-dissipative component of the current vanishes.  We focus on the regime of  low bias voltages, where the dissipative component of the current is dominated by the diagonal contribution.

We choose here to work in Eularian variables. To obtain the expression for the nonlinear conductance in this regime we substitute Eqs.~\eqref{eq:J_Lagr} and \eqref{nGenL} into Eq.~\eqref{eq:G_constant_U_def}.
It is convenient to change from integration over time $\tau$ in Eq.~\eqref{nGenL} to an integration over phase $\phi$,
\begin{align}
n(N,t) =&  \frac{1}{2eU\tau_{in}} \int^\infty_0 d\phi e^{-\phi/2eU\tau_{in}} n_F\big[ \epsilon \big( N, \chi(t) - \phi \big) \big].
\end{align}
When the temperature is large as compared to the typical range of motion of the quasi-particle energy levels,
we can expand the Fermi function deviations of the instantaneous quasi-particle energies from their average positions $\langle \epsilon(N,\phi ) \rangle_\phi$,
\begin{align}
\label{delta_epsilon_N_def}
 \delta \epsilon(N, \chi)  \equiv  &  \, \epsilon(N, \chi ) - \langle \epsilon(N,\phi ) \rangle_\phi.
\end{align}
This yields,
\begin{align}
\label{distributionEuler}
n\big(N, t \big) = & \,  n_F\big[ \langle \epsilon(N,\phi ) \rangle_\phi \big]    \nonumber \\
&  + \,  \frac{\partial_\epsilon n_F[ \langle \epsilon(N,\phi ) \rangle_\phi \big] }{2eU\tau_{in}} \int^\infty_0 d\phi  
\exp\left( - \frac{\phi}{2eU\tau_{in}}\right) 
\delta \epsilon(N, \chi(t) - \phi) .
\end{align}
Expanding the periodic phase dependence of the energy of quasi-particle levels in a Fourier series,
\begin{equation}
\label{fourier}
\delta \epsilon\big( N, \chi) =  \sum_{k\neq 0} C_k\big( N\big) e^{ik\chi},
\end{equation}
and using Eqs.~\eqref{distributionEuler},~\eqref{fourier} and the expression for the current Eq.~\eqref{EQ:CURRENTEUL},
we obtain the following  expression for the non-linear conductance
\begin{equation}\label{eq:cond_general}
\begin{split}
 \bar{G}_d(U)  =
 -4e^2 \tau_{in}  \int^\infty_0 dN  \partial_\epsilon n_F\big[\big \langle \epsilon \big(N,\chi \big)  \big\rangle_\chi \big]
\sum_{k \neq 0}\frac{k^2 }{1 + (2keU\tau_{in})^2} | C_k(N)|^2 .
\end{split}
\end{equation}
At small voltages, $eU\ll \tau_{in}^{-1}$, we obtain the linear conductance,
\begin{equation}\label{eq:cond_linear}
\begin{split}
 \bar{G}_d(0)  =
 -4e^2 \tau_{in}  \int^\infty_0 dN  \partial_\epsilon n_F\big[\big \langle \epsilon \big(N,\chi \big)  \big\rangle_\chi \big]
\sum_{k \neq 0}k^2  | C_k(N)|^2 .
\end{split}
\end{equation}
Comparing with Eq.~\eqref{eq:condChi_a} we see that the linear conductance
can be equivalently expressed in the terms of the phase dependent conductance $G_d[\chi]$ introduced in Eq.~\eqref{eq:G_chi_def},
\begin{align}
\label{eq:G_average_chi}
 \bar{G}_d(0) = \int_{0}^{2\pi} \frac{d \chi}{ 2\pi} G_{d} [\chi].
\end{align}
At large voltages,  $eU\gg  \tau^{-1}_{in}$,  Eq.~\eqref{eq:cond_general} yields
\begin{equation} \label{eq:conductance_large_U}
\bar{G}_d(U) =  -  \frac{1}{U^2\tau_{in}}\int^\infty_0 dN  \partial_\epsilon n_F\big[\big \langle \epsilon \big(N,\chi \big)  \big\rangle_\chi \big]
\sum_{k \neq 0} | C_k(N)|^2  .
\end{equation}
For a typical phase-dependence of the quasi-particle spectrum, the Fourier sums in Eqs.~\eqref{eq:cond_linear} and \eqref{eq:conductance_large_U} are dominated by $k$ of order unity. In this case, nonlinear conductance at  $eU\gg  \tau^{-1}_{in}$ can be estimated as
\begin{eqnarray} \label{GlargeU}
 \bar{G}_d(U)
 \sim  \frac{  \bar{G}_d(0) }{(eU \tau_{in})^{2}}.
 \end{eqnarray}

According to Eq.~\eqref{GlargeU}, at $eU\gg \tau^{-1}_{in} $ the \emph{dc} current $ \langle J\rangle =  \bar{G}_d(U)  U$ decreases as the voltage increases.
Thus, $\langle J\rangle$  has a maximum at $eU\sim  \tau^{-1}_{in} $.   
 The maximal current, 
 \begin{align}\label{eq:J_max_G_d}
 J_{max} \sim \frac{\bar{G}_{d}(0)}{e \tau_{in}},
 \end{align}
can be expressed in terms of the $\chi$-dependence of the quasi-particle spectrum using Eqs.~\eqref{eq:G_average_chi} and  \eqref{eq:G_chi_def}. In the Lagrangian and Eulerian variables the corresponding expressions have the form 
\begin{eqnarray}\label{eq:max_current}
J_{max}
&\sim  &  -4e  \int d \chi \sum_i \partial_\epsilon n_F(\epsilon_i) \big(\partial_\chi \epsilon_{i}(\chi(t)) \big)^{2} \nonumber   \\
 &=&  -4e  \int d \chi  \int_{0}^{\infty}  d \epsilon \nu(\epsilon, \chi) V^2_{\nu}(\epsilon, \chi)
\partial_\epsilon n_F(\epsilon).
\end{eqnarray}
It is worth noting that, since at high temperatures the equilibrium critical current $J_{c}(T)$ is exponentially decaying function of $T$, the value of $J_{max}$ can be much larger $J_{c}(T)$, and in some cases it can be as large as critical super-current at zero temperature $J_{c}(0)$.

Equation~\eqref{GlargeU} describing  the decrease of the nonlinear conductance with increasing voltage applies as long as the non-diagonal contribution to the dissipative current   $J_{nd}  = G_{nd} U(t) $ can be neglected.
At voltages 
\begin{equation}
U\sim U_{min}\equiv \frac{1}{\tau_{in}}\bigg[\frac{ \bar{G}_d(0) }{ G_{nd}}\bigg]^{1/2} \gg \frac{1}{\tau_{in}},
\end{equation}
the $I-U$ characteristic develops a minimum. At $U> U_{min}$ the dissipative current is dominated by the non-diagonal contribution $J_{nd}$,  which increases with $U$. It
has been studied in many articles, see for example Refs.~\cite{Klap,Bin1,Bin2,Naz,Klap4,Averin}. The shape of the $I-U$ characteristics of the junctions of voltage-biased junctions is illustrated in Fig.~\ref{FigfixedV_IVC.pdf}.
A somewhat different mechanism of N-type I-U characteristics of weak link has been  discussed in Refs. ~\cite{Volk, AslLark,Tinkh}.
 \begin{figure}[h]
\includegraphics[scale = 0.4]{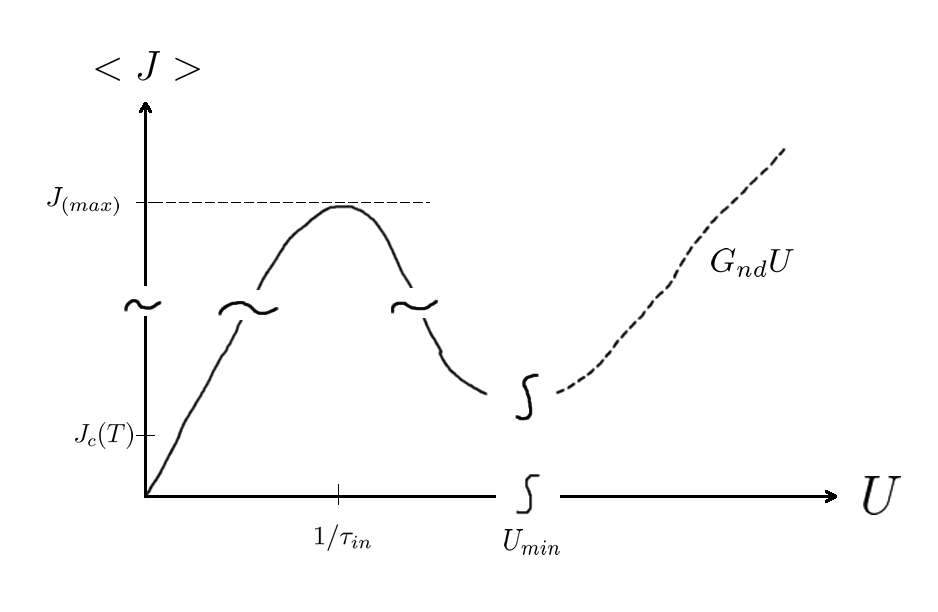}
\caption{A schematic picture of an I-U characteristics of a voltage-biased SNS junction. The value of $J_{(max)}>J_{c}(T)$  can be significantly larger than the value  of the equilibrium  critical current of the  junction $J_{c}(T)$. }
\label{FigfixedV_IVC.pdf}
\end{figure}

\subsection{I-U characteristics of current-biased junctions}
\label{sec:current-bias}

In the current bias setup, the SNS junction undergoes a transition into a resistive state when the bias current $J$ exceeds the critical current $J_c(T)$.
In this case the phase difference $\chi(t)$ increases monotonically, while the voltage $U(t)$ changes periodically with time, as illustrated in Fig.~\ref{Fig_Ut.pdf}.
In the following we will be interested in the dependence of the voltage averaged over the period of oscillations, $ \langle U(t) \rangle $,  on the bias current $J$.
Qualitatively, the I-U characteristics of the current-biased SNS junctions is shown in Fig.~\ref{IVC_fixedI.png}.
In a wide interval of bias currents $J > J_c (T)  $ the average voltage on the junction is relatively small because it is inversely proportional to the
the inelastic relaxation time $\tau_{in}$, which is the longest relaxation time in the system.  At a higher bias current, $J \approx J_{jump}$,  the voltage exhibits a relatively sharp jump to a much larger value.  The magnitude of $J_{jump}$ turns out to be of the same order as the maximal current $J_{max}$ in the voltage-bias case, which is given by Eq.~\eqref{eq:max_current}.

We will focus on the range of bias currents $J_c(T)<J<J_{jump}$,  in which the current is dominated by   the diagonal component $J_{d}$.  It is important to note however that according to Eqs.~\eqref{eq:J_Lagr} and \eqref{EQ:CURRENTEUL},  at the time-reversal invariant points $\chi = \pi n$, where $n$ is an integer,  the sensitivity of all quasi-particle levels with respect to the phase change vanishes.  As a result, $J_{d}$ vanishes at these points, and  in some intervals near these points the bias current must be carried by the non-diagonal contribution, $J_{nd}$.  Thus, the phase and time periods of the oscillations can be separated into two diagonal and two non-diagonal intervals,
$t_{p}=(t_{d,1} + t_{d,2} + t_{nd,1} + t_{nd,2})$, and    $2\pi=\chi_{d,1}+\chi_{nd,1}+\chi_{d,2}+\chi_{nd,2}$, in which the bias current is dominated by the diagonal,
$J_{d}$, or non-diagonal, $J_{nd}$, contributions respectively. The relatively sharp distinction between these two intervals is possible because $ \bar{G}_d(0)  \gg G_{nd}$.

The boundaries of the non-diagonal intervals  $\chi_{nd}$ can be determined from the condition that,  at $\dot{\chi} \sim 1/\tau_{in}$ the bias current can be carried by the maximal diagonal contribution, $J_d \sim G_d(\chi)/e\tau_{in} =J$.  In the vicinity of the time-reversal invariant points, $\chi = \pi n +\delta \chi$, we have
\begin{equation}
G_d(\chi) \sim \frac{\delta \chi^{2}}{2} \left. \frac{ d^2 G_{d}(\chi)}{d \chi^2}\right|_{\chi= \pi n} \sim \bar{G}_d(0) \frac{\delta \chi^{2}}{2}.
\end{equation}
As a result,  we get the following estimate for the width of the non-diagonal phase intervals: $\chi_{nd}  \sim  \sqrt{J/J_{jump}} $.

Inside the diagonal interval the phase winds at a rate of order of
$eU_{d}=eJ/\bar{G}_{d}(0)$, whereas inside the non-diagonal interval it winds at a rate
$eU_{nd}= eJ/G_{nd}$.
Therefore we can neglect   $t_{nd}\sim (G_{nd}/ \bar{G}_{d}(0))
(\chi_{nd}/\chi_{d}) t_{d}\ll t_{d}$  in Eq.~\eqref{eq:avg_voltage}. Thus, using  the Josephson relation  \eqref{eq:Joshepson}, the average voltage can be expressed in terms of the duration of the diagonal time intervals only,
\begin{equation}\label{eq:avg_voltage}
\langle U \rangle =  \frac{\pi}{e t_p}  \approx  \frac{\pi}{e \left(t_{d,1} + t_{d,2} \right)}.
\end{equation}
If the instantaneous phase is not to close to the time-reversal invariant points, $\chi =  n\pi $, the rate of change of phase is small, and the current may be expressed in terms of the instantaneous phase $\chi$  and its derivative via Eq.~\eqref{eq:G_chi_def}. 
Using this relation,  the duration of the diagonal time intervals may be expressed as
\begin{equation}\label{eq:td}
t_{d,i} = \frac{1}{2e} \int_{\chi_{d,i}} \frac{ G_{d}[\chi]  d\chi }{J - J_{c}(T)Y(\chi,T)},
\end{equation}
where $i=1,2$, and the integration  is taken over the phase interval $\chi_{d,i}$.

At small excess current, $J-J_{c}(T) \ll J_{c}(T)$, 
we can expand  $Y(\chi,  T)$ near its maximum at $\chi=\chi_{m}$, while at $J>J_{c}$ we can neglect the second term in the denominator in Eq.~\eqref{eq:td}.  Then, using  Eq.~\eqref{eq:avg_voltage} we get 
\begin{equation}\label{eq:avU}
\langle U(J) \rangle=
\begin{cases}
  \sqrt{2J_c (J - J_c)}/G_{d}[\chi_{m}] &  ,J-J_{c}  \ll J_{c} ,\\
\sim    J/  \bar{G}_{d}(0)  &  ,J_{c}\ll J\ll J_{jump}.
\end{cases}
\end{equation}
According to Eqs.~\eqref{eq:condChi_ab} and \eqref{eq:avU},  at relatively small currents the voltage across the junction  is smaller than  $1/\tau_{in}$.  This justifies  the use of linear in $\dot{\chi}$ approximation for the dissipative part of the current through the junction.

Similarly to the case of voltage-based junction, in the current-biased case
the diagonal component of the current $J_{d}$ has a maximum at $U\sim 1/e\tau_{in}$, which is of order of $J_{jump}$.
When the bias current, $J$ reaches this value, the widths of the  phase intervals $\chi_{d,i}$  shrinks to zero, and the voltage-current dependence  $\langle U(J) \rangle$  jumps to the branch dominated by the non-diagonal contribution to the current $J_{nd}$.

Near $J = J_{jump}$ the non-diagonal interval covers nearly the entire phase interval from $0$ to $2\pi$, with the exception of points near $\chi_{max}^{(G)}$, where $G_{d}[\chi]$ reaches its maximum. Expanding $G_{d}[\chi]$ near its maximum, we estimate the conductance at the edge of the diagonal interval
\begin{equation}
G_{d}[\chi_{max}^{(G)} + \chi_d] \sim J_{jump} \tau_{in} - \bar G_d(0) \chi^2_d.
\end{equation}
If the current is fixed at $J = J_{jump} - \delta J$, then the size of the diagonal interval is given by $\delta J \sim \bar G_d(0) \chi^2_d / \tau_{in} \sim J_{jump} \chi_d^2$. Within the width of the jump the diagonal and non-diagonal time intervals are of the same order $t_{nd} \sim t_{d}$, which means that $\chi_{d}  \sim \bar G_d(0) / G_{nd} $. Therefore, we estimate the width of the jump to be,
\begin{equation}
\delta J\sim J_{jump} \bigg( \frac{G_{nd}}{ \bar{G}_{d}(0)} \bigg)^2  \ll J_{jump}.
\end{equation}

In the regime $J>J_{jump}$ the voltage on the junction $U\sim J/G_{nd}\gg 1/\tau_{in}$, the diagonal contribution to the current $J_{d}$ is suppressed, and the  I-U characteristics of the junctions are controlled by the non-diagonal contribution to the current,
$
J_{nd}=G_{nd}U
$.

\begin{figure}[ptb]
\includegraphics[scale=0.34]{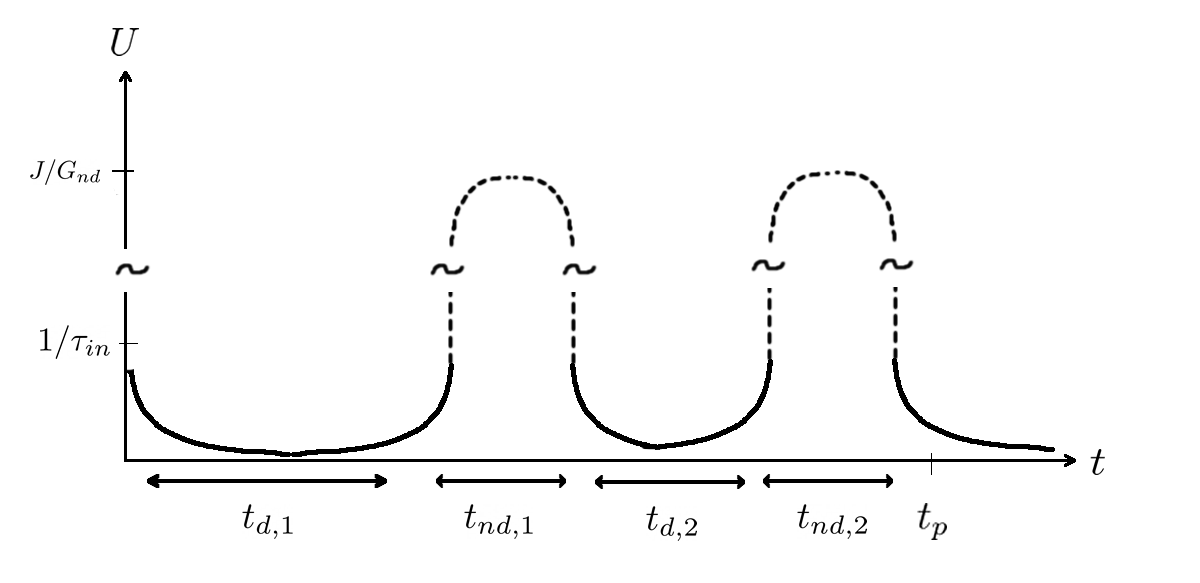}
\caption{Time dependence of  voltage at a current-based SNS junction when $J>J_{c}(T)$.
}
\label{Fig_Ut.pdf}
\end{figure}
\begin{figure}[h]
\includegraphics[scale = 0.4]{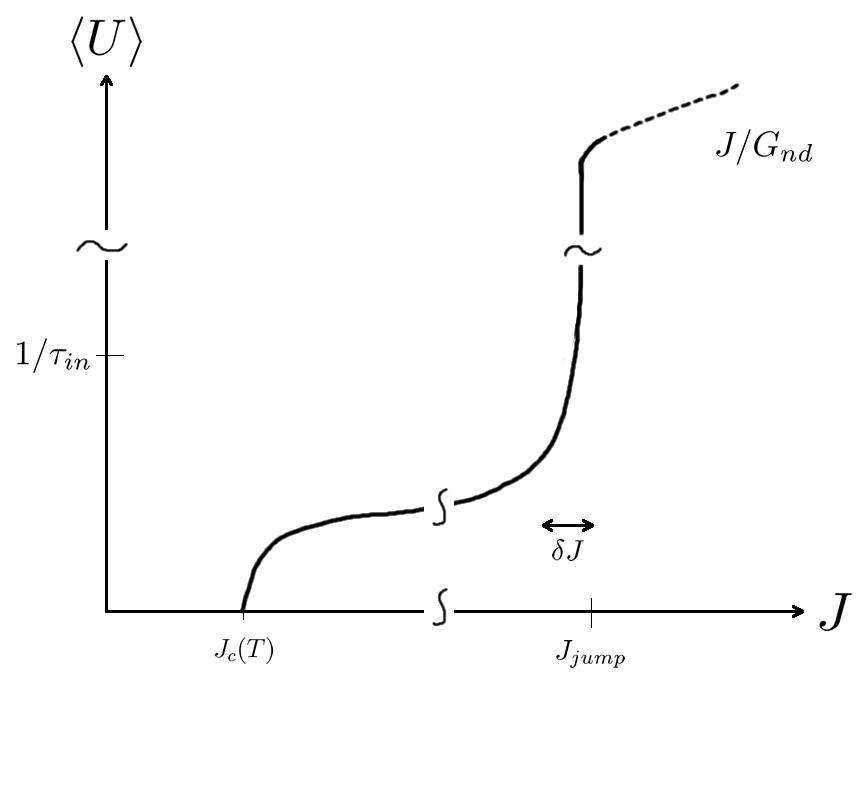}
\caption{ Schematic picture of the I-U characteristic of the current-based SNS junction.}
\label{IVC_fixedI.png}
\end{figure}

\section{I-U characteristics of SNS junctions in clean and diffusive regimes}\label{sec:Character}

As was shown in Sec.~\ref{sec:IU_general}, the I-U characteristics of both voltage-  and current- biased SNS junctions can be characterized by the parameter $ \bar{G}_{d}(0) \sim G_{d}[\chi_{m}]$; see  Eqs.~\eqref{eq:G_average_chi}, \eqref{GlargeU}, and \eqref{eq:avU}. In this section we will evaluate this parameter  in the cases of ballistic single channel junctions,  and diffusive multi-channel junctions.

\subsection{Clean 1D SNS junction.} \label{sec:clean}

In this subsection we consider a junction, in which the normal region consists of a clean single channel metallic wire. 
We assume that the length of the wire $L$ is    larger than the superconducting coherence length.  
In this case one can evaluate the quasi-particle spectrum by solving the stationary Bogoliubov-De Gennes equations in the normal metal at fixed value  of $\chi$ with appropriate boundary conditions at NS boundaries (see for example Refs.~\cite{Kulik} and \cite{Binn} )
 \begin {equation}\label{eq:BDG}
  \begin{bmatrix}
    \frac{\hat{p}^2}{2m}-\mu & 0 \\
      0 &  - \frac{\hat{p}^2}{2m}+\mu
  \end{bmatrix}  \begin{bmatrix}
   \psi_e\\
   \psi_h
  \end{bmatrix}= \epsilon \begin{bmatrix} \psi_e\\
                                                          \psi_h\end{bmatrix}.
\end{equation}
Below we assume that the transmission coefficients of both contacts are the same and equal to $r$. In the limiting cases of high and low transparency the spectrum for $\epsilon < \Delta$ is given by
\begin{equation}\label{eq:1dspectr}
\epsilon^{\pm}_{n} (\chi)= \frac{v_F}{L} \begin{cases}
\pi(n+\frac{1}{2}) \pm  \frac{\chi}{2},&  r = 1, \\
n \pi \pm 2  r \sqrt{2(1 + \cos\chi) }, & r \ll 1.
\end{cases}
 \end{equation}
Here $n=0,1...$ is integer, $v_{F}$ is the Fermi velocity, and the phase $\chi$ is understood modulo $2\pi$.

Below we evaluate the linear conductance $\bar{G}_d(0)$ given by  Eqs.~\eqref{eq:condChi_a}  and \eqref{eq:G_average_chi}, which can then be used to determine  the values of the maximal current  in the voltage-biased set up, $J_{max}$ and the current at which the transition to a high resistance state occurs in the current- biased set up, $J_{jump}$, via Eq.~\eqref{eq:max_current}.

Substituting Eq.~\eqref{eq:1dspectr} into Eq.~\eqref{eq:condChi_a} we  obtain an expression for the linear conductance  at high temperatures
\begin{equation}\label{clean G_0}
 \bar{G}_d(0) =\frac{e^2}{\pi } \frac{v_F}{L}\tau_{in}  A(r), \,\,\,\,\   T\gg v_{F}/L,
\end{equation}
where
\begin{equation}
A(r) =
\begin{cases}
1 &, r=1,\\
8r^2 &,  r \ll 1.
\end{cases}
\end{equation}
We note that the conductance of a pure single channel SNS junction, 
Eq.~\eqref{clean G_0},
exceeds the normal state conductance  $Ae^{2}/\hbar$,  by a large factor  $\frac{v_{F}}{L}\tau_{in}\gg 1$.
Substituting  Eq.~\eqref{clean G_0} into Eq.~\eqref{eq:max_current},  we get
\begin{align}
\label{eq:J_max_1D}
 J_{jump}\sim & J_{max} \sim  \frac{e v_F}{L} A(r).
\end{align}
The maximal current turns out to be temperature independent. The reason for this is that at low energies, $\epsilon_{n}\ll \Delta$, 
the sensitivity of the levels to a change in $\chi$ is independent of the energy.

 It is instructive to compare value of  $J_{max}$ and $J_{jump} $ in Eq.~\eqref{eq:J_max_1D} with  the critical current $J_{c}(T)$. The latter
 can be obtained by
substituting Eq.~\eqref{eq:1dspectr}, and the
equilibrium Fermi distribution function
$
n_{F}(\epsilon_{i})
$
 into Eq.~\eqref{eq:J_Lagr},   see Ref.~\cite{Kulik}.
\begin{equation}\label{eq:J_c_T_1D}
J_{c}(T) = B(r) \frac{ev_F}{2L } \begin{cases}
1 &, \quad T\ll \frac{v_{F}}{L},  \\
  \exp(- \frac{2 \pi T L}{v_F} ) &, \quad T\gg \frac{v_{F}}{L}.
  \end{cases} \end{equation}
where the dimensionless coefficient  $B(r)$ has the following limiting values at high and low contact transparencies,
\begin{equation}
B(r) =
\begin{cases}
1 &, r=1,\\
r^2/ 2\pi  &,  r \ll 1. 
\end{cases}
\end{equation}
Comparing Eqs.~\eqref{eq:J_max_1D}  and \eqref{eq:J_c_T_1D} we arrive to a somewhat surprising conclusion that at high temperatures, $T\gg v_{F}/L$, the values of $J_{max}$ and $J_{jump} $ are
 of order of the critical current at zero temperature,
\begin{equation}
 J_{max}\sim J_{jump}  \sim J_{c}(0)  \gg J_{c}(T).
\end{equation}

 At small temperatures, $T\ll v_{F}/L$,  the situation depends on the value of the transmission coefficient $r$. At $r=1$ the gap in the quasi-particle spectrum closes at $\chi=0,\pi$.  In this case the main contribution to     $ \bar{G}_d(0)$ 
comes from the interval of times where the gap is of order $T$.  In this case Eqs.~\eqref{eq:condChi_ab}, \eqref{eq:G_average_chi}, \eqref{eq:J_max_G_d}
 yield the  same value for $\bar{G}$, $J_{jump}$, and $J_{max}$  as in Eqs.~\eqref{clean G_0}, \eqref{eq:J_max_1D}.

 If $r<1$,  the gap in the spectrum does not close at any value of $\chi$. Therefore, at  $T\ll v_{F}/L$ quasi-particle concentration inside the junction is exponentially low $n_{i}\sim \exp(-v_{F}/LT)$.   In this case there are two relaxation times characterizing the dynamics of  the system:
 relaxation time characterizing processes which conserve the total number of quasi-particles, $\tau_{in}$ , and  the exponentially long  recombination time ,$\tau_{in,r}\sim \tau_{in} \exp(v_{F}/LT)$, which characterizes the processes changing the total number of particle. The two exponential factors are canceled in Eq.~\eqref{eq:J_Lagr} and we can estimate the low temperature linear conductance  to be roughly the same order as in the high temperature case,
\begin{equation}\label{clean G_0 2}
 \bar{G}_d(U)   \sim e^{2} \frac{v_F}{L} \tau_{in}\begin{cases}
1 &  ,(eU \tau_{in,r})\ll 1  ,\,\,\,\,\,\,\,   r = 1 ,\\
  (eU\tau_{in,r})^{-2}  & ,(eU \tau_{in,r})\gg 1 ,\,\,\,\,\,\,\,   r= 1.
\end{cases}
\end{equation}
Note that  since in this case $U_{max} \sim 1/e  \tau_{in,r}\ll 1/\tau_{in}$, the value of the maximum current in the adiabatic regime turns out to be smaller than its value at high temperatures by an exponentially small factor $ e^{-v_{F}/LT}$ .
Accordingly, at small temperatures, $T\ll v_{F}/L$,  we have  $J_{max} \ll J_{c}(0) $.

\subsection{Diffusive SNS  junctions.}
\label{diffusive case}

Let us now consider the case of a diffusive SNS junction shown in Fig.~\ref{SNSFIG_B} ,
 where two sides of a diffusive metal with the dimensions $L,L_{1},L_{2}\gg l=v_{F}\tau_{el}$ are attached to two superconducting parts of the junction, while the other two sides are in contact with insulator.

A general scheme of description of the kinetic phenomena in superconductors in the diffusive regime ($L\gg l$) has been developed by Larkin and Ovchinnikov~\cite{LarkinOvchinnikov}.
It describes both diagonal $J_{d}$ and non-diagonal $J_{nd}$ parts of the current as long as  $eU<\Delta$.
In the appendix we review derivation Larkin-Ovchinnikov equations and show that at $eU< E_{T}$ they can be reduced to Eqs.~\eqref{EQ:N_DOT} and \eqref{EQ:CURRENTEUL}.   Here $E_T = \frac{D}{L^2}$ is the Thouless energy and $D = v_F^2 \tau_{el} / 3$ is the diffusion coefficient in the normal metal.

The density of states in the normal metal part of the junction  can be written in terms of retarded Green's function,
\begin{equation}
\nu(\epsilon, \chi) =
\int_{V} d {\bf r} \tilde{\nu}(\epsilon, {\bf r}, \chi)={2} \tilde \nu_N  \mathrm{Re} \int_{V} d {\bf r} G_0^{R} (\epsilon, {\bf r}, \chi) . \color{black}
\end{equation}
Here the integral is taken over the normal metal region, $\tilde{\nu}({\bf r})$ is the local density of states in SNS junction,  and $\tilde{\nu}_N $ is the density of states per unit volume {per spin projection} of the normal state, and $G_0^{R}(\epsilon, {\bf r},\chi)$ is the dimensionless semi-classical retarded Green's function~\cite{LarkinOvchinnikov}.

In the geometry of the SNS junction shown in Fig.~\ref{SNSFIG}b, $\tilde{\nu}$ depends only on the $x$ coordinate.
Using the normalization condition for the normal and anomalous retarded Green's functions (see Eq.~\eqref{normalization gr2} in the appendix),  we parameterize them as follows
\begin{equation}
\label{usadel param}
G_0^{R}(\epsilon,x) = \cos\theta(\epsilon,x), \,\,\
F_0^{R} = e^{i\tilde{\chi}(\epsilon,x)} \sin\theta(\epsilon,x), \,\,\ F_0^{R+} = e^{-i\tilde {\chi}(\epsilon,x)} \sin\theta(\epsilon,x),
\end{equation}
and 
\begin{equation}
\label{eq:Uzadel2}
 \nu(\epsilon, \chi) ={ 2} \tilde \nu_N (L_1L_2) \int^\frac{L}{2} _{-\frac{L}{2}} dx  \mathrm{Re} \big[\cos\theta(\epsilon,x) \big ]. 
\end{equation}
Here $\theta$ and $\chi$ are complex.

In the diffusive regime   the dependence of  $\theta$  and $\tilde{\nu}(\epsilon, \chi, {\bf r})$ on $x$ and the phase difference $\chi$  can be obtained by solving the Usadel equations \cite{eq:uzadel} (see Eqs.~\eqref{usadel1} and \eqref{usadel2} in the Appendix).
In  the normal region, where $\Delta({\bf r})=0$ 
they have the form 
\begin{align}
\label{eq:Uzadel}
\frac{D}{2} \bigg(\partial^2_x \theta - \frac{1}{2} \sin(2\theta) \big(\partial_x \tilde{\chi} \big)^2 \bigg) = &
-\epsilon \sin\theta ,\\
\label{eq:Uzadel1}
\partial_x \big(\sin^2\theta \partial_x \tilde \chi \big)= &  \, 0.
\end{align}

The boundary conditions for these equations  at  $\epsilon<\Delta$ and  $r =1$  are  (see Ref.~\cite{Kurpianov})
\begin{equation}
\begin{split}
\theta(\epsilon; x=0,L ) = \frac{\pi}{2}  ; \,\,\,\,\,\,
\tilde \chi(\epsilon;x=0,L ) = \pm \frac{\chi}{2}.
\end{split}
\end{equation}
For $r\ll 1$ the boundary conditions have a form
\begin{equation}
\begin{split}
D\partial_x \theta|_{\epsilon; x = 0,L}= \pm r v_F \big( \cos \theta \big) \cos(\tilde \chi \pm \frac{\chi}{2}) |_{\epsilon; x = 0,L} ,  \\
D\sin\theta \partial_x \tilde \chi |_{\epsilon; x = 0,L}= \pm  r v_F \sin(\tilde \chi \pm \frac{\chi}{2})|_{\epsilon; x = 0,L}.
\end{split}
\end{equation}
Solutions of Eqs.~\eqref{eq:Uzadel},\eqref{eq:Uzadel1}  were investigated in several articles (see for example,  Refs.~\cite{Naz} and \cite{Pannetier}).

The density of states in the normal  region of  SNS junctions differs from that in the normal metal only at small energies of the order of mini-gap $E_{g}$.
For our purposes we need only rough features of  $\epsilon$ and $\chi$ dependencies of the density of states,
\begin{equation}
\label{eq:DensityOfStates}
\nu(\epsilon, \chi) =  {2} v\tilde \nu_N \begin{cases}
0 & ,\epsilon < E_g(\chi),\\
h(\epsilon, \chi)  & ,\epsilon - E_g(\chi) \sim E_g(\chi), \\
1 & ,\epsilon - E_g(\chi) >> E_g(\chi).
\end{cases}
\end{equation}
where $h(\epsilon, \chi)$ is of order unity,  and $v=LL_{1}L_{2}$ is the volume of the normal metal region.
When the phase winds from $0$ to $2\pi$ the value of mini-gap changes on the order of
\begin{equation}
E_{g}(0)\sim
E_{T} \begin{cases}
 1  &, r> \frac{l}{L}, \\
 r\frac{L}{l}  & ,r< \frac{l}{L},
\end{cases}
\end{equation}
which implies that $\partial_\chi E_g(\chi) \sim E_{g}(0)$.

Substituting Eq.~\eqref{eq:DensityOfStates} into Eqs.~\eqref{eq:level_velocity},\eqref{eq:condChi_ab}
  and averaging the result over the period of oscillations we can estimate the conductance of the junction  as follows
\begin{equation}
\label{eq:GaverageD}
 \bar{G}_d(0,T)  \sim G'_{N}  \tau_{in}  \begin{cases}
\frac{E_{g}^{2}(0)}{T}&   \,\,\,\ ,T>E_{g}(0), \,\,\,\,\   U\ll 1/\tau_{in}, \\
T  &  \,\,\,\,\, ,T<E_{g}(0),   \,\,\,\,\ U\ll 1/\tau_{in},   \\
\end{cases}
\end{equation}
where  $G'_{N} = e^2 E_{g}(0) \tilde \nu_{N} v$.  The situation at large voltages, $eU\gg \tau_{in}^{-1}$,  is similar to that described in Sec \eqref{sec:clean}  for a clean one-dimensional SNS junction.   Namely,  the nonlinear conductance $ \bar{G}_d(U)$ is reduced from its linear value,  Eq.~\eqref{eq:GaverageD},  by the factor $(2eU\tau_{in})^{-2}$.
Thus,  the I-U characteristics of a voltage-biased junction has a maximum at $eU\sim \tau_{in}^{-1}$.  The magnitude of the maximal current can be estimated as
\begin{equation}
\label{eq:JmaxAv}
J _{max} \sim J_{c}(0) \begin{cases}
\frac{E_{g}(0)}{T} &   \,\,\,\ ,T>E_{g}(0), \,\,\,\,\    \\
\frac{T}{E_{g}(0)} &  \,\,\,\,\, ,T<E_{g}(0) .\,\,\,\,\      \\
\end{cases}
\end{equation}
Here $J_{c}(0)=\frac{1}{e}G'_{N}E_{g}(0)$  is the  critical current of a diffusive SNS junction at $T=0$.  We note that the value of $J_{max} $ can be significantly larger than $J_{c}(T)$.

At even larger larger voltages  the dominant contribution to the current comes from $J_{nd}$,  which is an increasing function of voltage.  Let us consider the case $T\gtrsim E_{g}$.  In this regime the part of the resistance  of the junction corresponding to $J_{nd}$ is,  essentially,   the resistance of the sequence of the tunneling barriers and the normal metal resistances. It has been considered in many articles \cite{Klap,Bin1,Bin2,Naz,Klap4}, \cite{Zhou} and \cite{Keles}.
For example,   if $r=1$ and  $E_{g}(0) \sim E_{T}$, then  the contribution to the current from the non-diagonal part  is on the order of the current in the normal state $J=G_{N}U$,  where $G_{N}=e^{2} E_{T}\tilde{\nu}_{N} v$  is the conductance of normal metal part of SNS junction    (see Eqs. \eqref{nd current}, \eqref{nd current 2} in the appendix).   As a result,  using Eq.~\eqref{eq:GaverageD},  we get an estimate for 
 \begin{equation}
 U_{min}\sim \frac{ E_{T}}{(T\tau_{in})^{1/2}}\ll E_{T}.
 \end{equation}

\section{Conclusions.} \label{sec:conclusions}

We have shown that   the I-U characteristics of SNS junctions at low temperatures and low voltages can be expressed in terms of the energy and the phase dependence of the density of states $\nu(\epsilon, \chi)$.
In this case, they are controlled by the inelastic  quasi-particle  relaxation time $\tau_{in}$. In contrast, 
at large bias voltages and currents the I-U characteristics are controlled by the elastic relaxation time $\tau_{el}$ .
 Qualitatively,  our results are shown in Figs.~\ref{FigfixedV_IVC.pdf} and \ref{IVC_fixedI.png}.

An interesting aspect of the problem is that for current-biased  junctions,  the jump in the I-U characteristics  from the low voltage to high voltage regime (see Fig.~\ref{IVC_fixedI.png})
occurs at the  value of the current $J=J_{jump}$,  which can be significantly larger then the value of the equilibrium critical current $J_{c}(T)$.   Therefore,  determination of the critical
 currents of  SNS junctions may require measurements of I-U characteristics at relatively small voltages,  $eU< 1/\tau_{in}$.

The results presented above are valid in situations where the low energy quasi-particles  are trapped inside the normal region of the junction,  and the only channel of the quasi-particle relaxation is the inelastic energy relaxation.  In a different geometry,  where  the normal region of the junction is open to the bulk normal metal,  as shown in  Fig.~\ref{SNSFIG_C},  there is another channel of the relaxation via diffusion of quasi-particles into the bulk of the normal metal.  In this case one can  obtain  an estimate for the conductance of the system substituting in  Eq.~\eqref{eq:GaverageD}
  \begin{equation}
  \tau_{in}\rightarrow min[\tau^{(*)}, \tau_{in} ]
 \end{equation}
where  $\tau^{*}\sim L^{2}_{1}/D$ is the time of diffusion on the length $L_{1}$.

  Finally, it should be mentioned  that the only symmetry requirement for the density of state in the time reversal symmetrical  system is $\nu(\epsilon, \chi, {\bf H})= \nu(\epsilon, -\chi, -{\bf H})$.   
  Therefore, for example,  in the case on non-centrosymmetric films in the parallel magnetic field $\nu(\epsilon, \chi, H)\neq \nu(\epsilon,\chi, -H)$ and $\nu(\epsilon, \chi, H)\neq \nu(\epsilon,-\chi, H)$.  As a result, in general,  the I-U characteristics of the SNS junctions are non-reciprocal:  $|J(U)\neq |J(-U)|$,  and $|J(H)\neq |J(-H)|$.

\renewcommand{\theequation}{A\arabic{equation}}

\section{Acknowledgment}
This work  of  was supported by the National Science Foundation Grant MRSEC DMR-1719797 and also in part by  the Thouless Institute for Quantum Matter and the College of Arts and Sciences at the University of Washington.  The work BZS was funded  by the Gordon and Betty Moore Foundation’s EPiQS Initiative through Grant GBMF8686.    

\appendix 

\section{Derivation of  Eqs.~\eqref{EQ:N_DOT},~\eqref{EQ:CURRENTEUL} using Larkin-Ovchinnikov approach in the diffusive regime, $L\gg l$.}
\label{sec:derivationLO}

\setcounter{equation}{0}

We start with the Gorkov equations for the Green's functions in Keldysh  representation \cite{Keldysh}.
We will denote matrices  in Nambu space with a hat, $\hat A$, and matrices in both Nambu and  Keldysh space with a check, $\check A$. We have chosen units such that $\hbar = c = 1$.
The Green's function is defined by the following equation, 
\begin{equation}
\begin{split}
\label{gorkov}
 \bigg( i  \hat \tau_3 \partial_{t_1}  + \frac{1}{2m} \bigg({\bf \nabla_{r_1}} - ie\hat \tau_3 {\bf A}({\bf r_1},t_1) \bigg )^2+ \mu \\+  \hat \Delta({\bf r_1},  t_1) - e \phi({\bf r_1},t_1)   \bigg)\check G ({\bf r_1}, {\bf  r_2},t_1, t_2)\\
  - (\check \Sigma \otimes \check G)({\bf r_1}, {\bf  r_2},t_1, t_2)    = \delta (\boldsymbol r_1 - \boldsymbol r_2) \delta (t_1 - t_2).
\end{split}
\end{equation}
Here $ {\bf A} ({\bf r},t)$ is the vector potential, $\hat \Delta({\bf r},t) $ is the superconducting order parameter, $\mu$ is the chemical potential,  and $\phi({\bf r})$ is the scalar potential,
\begin{align}
\check G =
\begin{pmatrix}
\hat G^R & \hat G^K \\
0 & \hat G^A
\end{pmatrix}, &&
\check \Sigma =
\begin{pmatrix}
\hat \Sigma^R & \hat \Sigma^K \\
0 & \hat \Sigma^A
\end{pmatrix},
\end{align}
where $\hat{G}^{R,A,K}$ are retarded, advanced, and Keldysh Green's functions, and $\check{\Sigma}$ is the self energy.
The cross operator in Eq.~\eqref{gorkov} represents a convolution,
\begin{equation}
 (O_1  \otimes O_2) ({\bf r_1}, {\bf r_2}, t_1, t_2) = \int d {\bf r} \int dt O_1 ({\bf r_1}, {\bf r}, t_1, t)  O_2 ({\bf r}, {\bf r_2}, t, t_2).
\end{equation}
The conjugate equation to Eq. \eqref{gorkov} is given by
\begin{equation}
\begin{split}
\label{gorkov2}
\check G ({\bf r_1}, {\bf  r_2},t_1, t_2)   \bigg( -i  \hat \tau_3 \partial_{t_2} + \frac{1}{2m} \bigg({\bf \nabla_{r_2}} + ie\hat \tau_3  {\bf A}({\bf r_2},t_2) \bigg)^2 +\\ \mu +  \hat  \Delta({\bf r_2},  t_2) - e \phi({\bf r_2},t_2) \bigg)\\
 - (\check G\otimes \check \Sigma)({\bf r_1}, {\bf  r_2},t_1, t_2)  = \delta (\boldsymbol r_1 - \boldsymbol r_2) \delta (t_1 - t_2),
\end{split}
\end{equation}
where the derivatives are understood to be acting towards the left.
These equations should be supplemented with the self-consistent equation for the order parameter
\begin{equation}
\Delta(r,t)  = \lambda  F(r,t),
\end{equation}
 where $\lambda$ is the electron interaction constant.

\subsection{ Quasi-classical approximation for  the Gorkov equations.}

Subtracting Eq.~\eqref{gorkov2} from Eq.~\eqref{gorkov} gives the following equation for the Green's functions,
\small
\begin{equation}
\begin{split}
\label{gorkov3}
 i  \hat \tau_3 \partial_{t_1}\check G({\bf r_1}, {\bf  r_2},t_1, t_2)  +  i\hat  \partial_{t_2} \check G({\bf r_1}, {\bf  r_2},t_1, t_2) \hat \tau_3 \\
 +\bigg (\frac{1}{2m} \bigg ({\bf \nabla_{r_1}} - ie {\bf A}({\bf r_1},t_1) \bigg )^2 + \mu + \hat \Delta({\bf r_1},  t_1) - e \phi({\bf r_1},t_1) \bigg) \check G({\bf r_1}, {\bf  r_2},t_1, t_2) \\
 -  \check G({\bf r_1}, {\bf  r_2},t_1, t_2) \bigg( \frac{1}{2m}({\bf \nabla_{r_2}} + ie \hat \tau_3 {\bf A}({\bf r_2},t_2) )^2 + \mu +  \hat  \Delta({\bf r_2},  t_2) - e \phi({\bf r_2},t_2) \bigg)    \\
  =   ( \check \Sigma \otimes  \check G) ({\bf r_1}, {\bf  r_2},t_1, t_2) -(\check G \otimes  \check \Sigma) ({\bf r_1}, {\bf  r_2},t_1, t_2) .
\end{split}
\end{equation}
\normalsize
In the limit where fields are slowly varying in space and time, we can use the quasi-classical approximation.
Introducing the Wigner coordinates,
\begin{align}
\begin{aligned}
\boldsymbol r = \frac{1}{2} (\boldsymbol r_1 +  \boldsymbol r_2), && \boldsymbol  {\tilde r} = \boldsymbol r_1 - \boldsymbol r_2 ,\\
t = \frac{1}{2} (t_1 +  t_2), && \tilde t = t_1 - t_2 ,
\end{aligned}
\end{align}
Fourier transforming equation Eq.~\eqref {gorkov3} over the relative position ${\bf \tilde r}$ as well as the relative time $\tilde t$, and dropping terms which are second order in derivatives,  we arrive at the following equation
\small
\begin{equation}
\label{quasi1}
\begin{split}
 \frac{1}{2} \partial_t \big \{ \hat \tau_3, \check G(\epsilon, {\bf r},t,{\bf p})  \big \} - i\epsilon \big [ \hat \tau_3, \check G(\epsilon, {\bf r},t,{\bf p})  \big ]
+ \frac{{\bf p} }{m}  \cdot {\bf \nabla_r} \check G(\epsilon, {\bf r}, {\bf p})\\
+ \big[\hat H({\bf r}, t, {\bf p}) , \check G(\epsilon,{\bf r}, t, {\bf p}) \big] 
- \frac{i}{2} \big \{ \partial_t \hat H({\bf r}, t, {\bf p}) , \partial_\epsilon \check G(\epsilon,{\bf r}, t, {\bf p}) \big \}\\
- \frac{e}{2m} {\bf A}({\bf r},t) \cdot {\bf \nabla_r} \big \{ \hat \tau_3, \check G(\epsilon, {\bf r}, t, {\bf p}) \big \}
+ \frac{i}{2} \big \{ {\bf \nabla_r} \hat H({\bf r}, t, {\bf p}), {\bf \nabla_p} \check G(\epsilon,{\bf r}, t, {\bf p}) \big \}\\
= -i\big [ \check \Sigma(\epsilon,{\bf r}, t, {\bf p}), \check G(\epsilon,{\bf r}, t, {\bf p}) \big]\\
+ \frac{1}{2} \big \{ {\bf \nabla_r}\check \Sigma (\epsilon,{\bf r},t,{\bf p}), {\bf \nabla_p}\check G (\epsilon,{\bf r},t,{\bf p}) \big \}
- \frac{1}{2} \big \{ {\bf \nabla_p}\check \Sigma (\epsilon,{\bf r},t,{\bf p}), {\bf \nabla_r}\check G (\epsilon,{\bf r},t,{\bf p}) \big \}\\
- \frac{1}{2}\big \{ \partial_t \check \Sigma (\epsilon,{\bf r},t,{\bf p}), \partial_\epsilon\check G (\epsilon,{\bf r},t,{\bf p}) \big \}
+ \frac{1}{2} \big \{ \partial_\epsilon \check \Sigma (\epsilon,{\bf r},t,{\bf p}), \partial_t \check G (\epsilon,{\bf r},t,{\bf p}) \big \}.
\end{split}
\end{equation}
\normalsize
Here the brackets $[\cdot, \cdot]$  and $\{\cdot , \cdot \}$ stand for commutators and anti-commutators, and we have defined,
\begin{align}
\check G(\epsilon, {\bf r}, t, {\bf p}) = \int dt \int d^3{\bf r} \check G({\bf r_1}, {\bf r_2}, t_1, t_2) e^{-i {\bf p}\cdot{\bf \tilde r} + i\epsilon \tilde t },\\
\hat H({\bf r}, t, {\bf p}) = \frac{-ie}{m} {\bf A}({\bf r}, t)\cdot {\bf p} \hat \tau_3  -i \hat \Delta({\bf r},t)
 + \frac{ie^2}{m} {\bf A}^2({\bf r}, t) + ie \phi({\bf r},t).&
\end{align}
\subsection{ The diffusion approximation for Gorkov equations.}

The self-energy $\check \Sigma=\check \Sigma_{el}+\check \Sigma_{in}$ is a sum of two contributions corresponding to elastic and inelastic scattering respectively.
In the case when the total scattering rate $\check \Sigma$ is smaller than the characteristic quasi-particle energy, it  can be dropped from the equation for the retarded Green's function.
In this case the quasi-particle momentum is a good quantum number, and one can use a conventional Boltzmann kinetic equation for quasi-particle distribution function to describe slow superconducting dynamics \cite{Aronov}.
In this article we will be interested  in the opposite limit, where the quasi-particle momentum is not a good quantum number, and
 \begin{equation}
\tau_{el}^{-1}>E_{T}>\tau_{in}^{-1}.  \end{equation}

We note that the Thouless energy, $E_{T}$ is a characteristic quasi-particle energy relevant to the problem.
In this case $\check{\Sigma}_{in}$  still can be dropped from the equation for the retarded Green's function, however $\check \Sigma_{el}$
is the largest term in Eq.~\eqref{quasi1}, and can not be neglected.

An effective approach to describe the quasi-particle dynamics in this limit was developed in Ref.\cite{LarkinOvchinnikov}.
This method is based on the fact that the elastic part of the self-energy can be expressed in terms of the Green's functions,
\begin{equation}
\label{self energy}
\check \Sigma_{el}(\epsilon, {\bf r}, t) = \frac{-1}{2\pi \tau_{el}} \int d^3 {\bf p} \check G(\epsilon, {\bf r},t, {\bf p}),
\end{equation}
and thus $\check \Sigma_{el}$ does not depend on  ${\bf p}$.

Let us integrate Eq.~\eqref{quasi1} over $\xi_p = \frac{{\bf p}^2}{2m} - \mu$ for a fixed momentum direction $\bm{n} = \bm{p}/p$.
On length scales larger then the Fermi wave length $p_F^{-1}$,  to leading order in spacial gradients we get
\begin{equation}
\label{quasigorkov}
\begin{split}
 \frac{1}{2} \partial_t \big \{ \hat \tau_3, \check g(\epsilon, {\bf r},t,{\bf n})  \big \} - i\epsilon \big [ \hat \tau_3, \check g(\epsilon, {\bf r},t,{\bf n})  \big ]
+ v_F {\bf n}  \cdot {\bf \nabla_r} \check g(\epsilon, {\bf r}, {\bf n}) \\
+ \big[\hat H({\bf r}, t, p_F {\bf n}) , \check g(\epsilon,{\bf r}, t, {\bf n}) \big]
- \frac{i}{2} \big \{ \partial_t \hat H({\bf r}, t, p_F {\bf n}) , \partial_\epsilon \check g(\epsilon,{\bf r}, t, {\bf n}) \big \}\\
= -i \big [\check \Sigma_{el}(\epsilon, {\bf r},t),  \check g (\epsilon,{\bf r}, t, {\bf n}) \big]-i \big [\check \Sigma_{in}(\epsilon, {\bf r},t),  \check g (\epsilon,{\bf r}, t, {\bf n}) \big],
\end{split}
\end{equation}
where we have defined
\begin{equation}
\check g(\epsilon,{\bf r}, t, {\bf n}) = \frac{i}{\pi} \int d\xi_p \check G(\epsilon, {\bf r}, t, {\bf p}).
\end{equation}
We have introduced the factor $i/\pi$ to have the same notation as in Ref.~\cite{LarkinOvchinnikov}.

Taking into account the normalization condition(see for example \cite{shelankov}),
\begin{equation}
\label{norm}
(\check g \otimes \check g) ({\bf r}, t_1,t_2, \boldsymbol n) = \delta (t_1 - t_2),
\end{equation}
we can parameterize the Keldysh component of $\check{g}$ as,
\begin{equation}
\label{gk}
\hat g^K(t_1,t_2, \boldsymbol n) = (\hat g^R\otimes \hat f) (r, t_1,t_2, \boldsymbol n) - (\hat f \otimes \hat g^A) (r, t_1,t_2, \boldsymbol n) .
\end{equation}
Since the matrix in the Nambu space $\hat f(r,t_1,t_2, {\bf n})$ has no off-diagonal component, we can expand it as
\begin{equation}
\hat f ({\bf r}, t_1,t_2, \boldsymbol n) =f({\bf r},t_1,t_2, \boldsymbol n) + \hat \tau_3 f_{1} ({\bf r},t_1,t_2, \boldsymbol n).
\end{equation}
To obtain $\hat g^K(\epsilon,{\bf r},t,{\bf n})$, we must Fourier transform Eq.~\eqref{gk} with respect to the relative time difference.
To zeroth order in time derivatives, we have
\begin{equation}
\hat g^K(\epsilon,{\bf r}, t, {\bf n}) =  \hat g^R (\epsilon,{\bf r}, t, {\bf n}) \hat f(\epsilon, {\bf r}, t,{\bf n}) - \hat f(\epsilon,{\bf r}, t, {\bf n})  \hat g^A (\epsilon,{\bf r}, t,{\bf n}).
\end{equation}
We can write $\hat{g}^K$ in the form,
\begin{equation}
\label{gkparam}
\hat g^K(\epsilon,{\bf r},t, {\bf n}) = 2 f(\epsilon, {\bf r},t, {\bf n}) \hat \delta(\epsilon,{\bf r},t, {\bf n})  + 2 f_{1}(\epsilon,{\bf r},t, {\bf n})\hat \alpha (\epsilon,{\bf r},t,{\bf n}),
\end{equation}
where we have defined,
\begin{equation}
2 \hat \alpha(\epsilon,{\bf r},{\bf n}) = \hat g^R(\epsilon,{\bf r},{\bf n}) \hat \tau_3 - \hat \tau_3 \hat g^A(\epsilon,{\bf r},{\bf n}),
\end{equation}
\begin{equation}
2 \hat \delta(\epsilon,{\bf r},{\bf n})  = \hat g^R(\epsilon,{\bf r},{\bf n})   - \hat g^A(\epsilon,{\bf r},{\bf n}).
\end{equation}

In the diffusive limit, where $\tau_{el}^{-1}$ is much larger than the typical energy scales of the problem, Greens functions are almost isotropic, and we can expand them in the spherical harmonics.
\begin{equation}
\label{spherical expansion}
\check g(\epsilon,{\bf r}, t, {\bf n}) = \check g_0(\epsilon,{\bf r}, t)  + \check {\bf g_1} (\epsilon,{\bf r}, t)  \cdot {\bf n},
\,\,\,\,\,\,\,\,\,\,\,\,\, \check  g_0(\boldsymbol r, t_1, t_2) \gg \check {\bf g}_1(\boldsymbol r, t_1, t_2) \cdot\boldsymbol  n.
\end{equation}
It follows from  the normalization condition Eq.~\eqref{norm}, that
\begin{equation}
\label{g0}
\check g_0(\epsilon,{\bf r}, t)  \check g_0(\epsilon,{\bf r}, t)  = 1,  \,\,\,\,\,\,\,\,\,\,\,\,\, \
\check {\bf g_1} (\epsilon,{\bf r}, t)  \check g_0(\epsilon,{\bf r}, t)  = -\check g_0(\epsilon,{\bf r}, t)  \check {\bf g_1}(\epsilon,{\bf r}, t).
\end{equation}
Substituting Eq.~\eqref{spherical expansion} into~\eqref{quasigorkov}, using Eq.~\eqref{g0} and the fact that
$\label{self energy} \check \Sigma_{el} = \frac{-i}{2\tau_{el}} \check g_0$,
in the linear in spacial gradients approximation we get
\begin{align}
\label{g1 relation}
\check {\bf g_1} (\epsilon,{\bf r}, t) =
&-\frac{3D}{v_F} \bigg(
\check g_0(\epsilon,{\bf r}, t) {\boldsymbol \partial_{\bf r}}  \check g_0(\epsilon,{\bf r}, t)\\ \nonumber&
\, + \frac{e}{2}\partial_t {\bf A}({\bf r}, t) \check g_0(\epsilon, {\bf r}, t)\big\{\hat \tau_3,\partial_\epsilon \check g_0(\epsilon, {\bf r}, t) \big \} \bigg) .
\end{align} \normalsize
Here  $\boldsymbol \partial_{\bf r} = {\bf \nabla_r} - ie {\bf A}({\bf r},t)[\hat \tau_3, \cdot ]$ is the covariant derivative. Substituting Eqs.~\eqref{spherical expansion},~\eqref{g1 relation} into~\eqref{quasigorkov}, and averaging the result over direction of ${\bf n}$, we get an equation for the isotropic part of the Green's functions $\check{g}_{0}$,
\small
\begin{equation}
\label{green diffusive}
\begin{split}
 \frac{1}{2} \partial_t \big \{ \hat \tau_3, \check g_0(\epsilon, {\bf r},t)  \big \} 
- i\epsilon \big [ \hat \tau_3, \check g_0(\epsilon, {\bf r},t)  \big ]
-D {\boldsymbol \partial_{\bf r}} \cdot\bigg(
\check g_0(\epsilon,{\bf r}, t)  {\boldsymbol \partial_{\bf r}} \check g_0(\epsilon,{\bf r}, t) \bigg)\\
+ \frac{eD}{2}\partial_t {\bf A} ({\bf r},t)  \partial_\epsilon \big\{ \hat \tau_3 ,  \check g_0 (\epsilon, {\bf r}, t) {\boldsymbol \partial_{\bf r}}  \check g_0 (\epsilon, {\bf r}, t) \big\}
-i\big[ \hat \Delta({\bf r},t),  \check g_0 (\epsilon, {\bf r}, t) ]\\ - \frac{1}{2} \big \{ \partial_t \hat \Delta({\bf r},t),  \partial_\epsilon \check g_0 (\epsilon, {\bf r}, t) \big \}
+ \partial_t \phi({\bf r},t)  \partial_\epsilon \check g_0 (\epsilon, {\bf r}, t)\\
= -i \big [\check \Sigma_{in}(\epsilon, {\bf r},t),  \check g_0 (\epsilon,{\bf r}, t) \big].
\end{split}
\end{equation}\normalsize

In the adiabatic approximation, valid when the external perturbations vary slowly compared to $\Delta^{-1}$ , time derivatives can be dropped in the diagonal components of
Eq.~\eqref{green diffusive} and we get Usadel's equations \cite{eq:uzadel} in matrix form
\begin{equation}
\label{usadel greens}
\begin{split}
i\epsilon \big [ \hat \tau_3, \hat g^R_0(\epsilon, {\bf r},t)  \big ]
+D {\boldsymbol \partial_{\bf r} } \cdot\bigg(
\hat g^R_0(\epsilon,{\bf r}, t)  \boldsymbol \partial_{\bf r} \hat g^R_0(\epsilon,{\bf r}, t) \bigg)
\\ + i \big [ \hat \Delta({\bf r}, t), \hat g^R_0(\epsilon, {\bf r}, t) \big]
= 0.
\end{split}
\end{equation} 

To get the two equations $f$ and $f_1$, we look at the Keldysh component of Eq.~\eqref{green diffusive} and take the trace in Nambu space (multiplying by a factor of $\hat \tau_3$ before taking the trace to get the second equation). As a result we have,
\begin{equation}
\label{diffusive f}
\begin{split}
\tilde \nu(\epsilon, {\bf r},t) \partial_t  f(\epsilon, {\bf r},t) + \frac{1}{4}e D \tilde \nu_N \partial_\epsilon f(\epsilon, {\bf r},t)  \bigg( \partial_t {\bf A} ({\bf r},t) \cdot {\bf j_\epsilon}({\bf r},t)
\\- 2\mathrm{Tr} \big \{ \partial_t \hat \Delta({\bf r},t) \hat \delta(\epsilon, {\bf r},t) \big\}\bigg) 
- \frac{1}{4} D\tilde \nu_N {\bf \nabla_r}  \cdot \big(    \Pi_1(\epsilon, {\bf r},t) {\bf \nabla_r} f(\epsilon, {\bf r},t)\big)\\
- \frac{1}{4}  D\tilde \nu_N {\bf \nabla_r} \cdot \big( f_1(\epsilon, {\bf r},t){\bf j_\epsilon}({\bf r},t) \big)
= I_1 \{f \},
\end{split}
\end{equation}
\begin{equation}
\label{diffusive f1}
\begin{split}
\partial_t \big(f_1(\epsilon, {\bf r},t) \tilde \nu(\epsilon, {\bf r},t) \big)
-  \frac{1}{4} D \tilde \nu_N {\bf \nabla_r}\cdot \big( \Pi_2(\epsilon, {\bf r},t)  {\bf \nabla_r} f_1(\epsilon, {\bf r},t)  \big) \\
- D\tilde \nu_N {\bf \nabla_r} \cdot \big( f(\epsilon, {\bf r},t) j_\epsilon({\bf r},t) \big)
- \frac{i}{2} f_1(\epsilon, {\bf r},t) \mathrm{Tr} \big \{\big (g^R(\epsilon, {\bf r},t)  \\+ g^A(\epsilon, {\bf r},t) \big)\hat \Delta(\epsilon, {\bf r},t) \big\}
+ \frac{1}{2} \partial_\epsilon f(\epsilon, {\bf r},t) \mathrm{Tr} \big \{ e \partial_t \phi \hat \alpha(\epsilon, {\bf r},t)
 - \\\partial_t \hat \Delta(\epsilon, {\bf r},t) \big\}
= I_2 \{f_1 \}.
\end{split}
\end{equation}
Here $\mathrm{Tr} $ stands for a trace in the Nambu space, and we have defined,
\begin{equation}
\Pi_1(\epsilon, {\bf r},t)  = \mathrm{Tr}\{ 1- \hat g^A_0 \hat g^R_0 \},
\end{equation}
\begin{equation}
\Pi_2 (\epsilon,{\bf r}, t) = \mathrm{Tr}\{ 1- \hat \tau_3 \hat g^A_0 \hat \tau_3\hat g^R_0 \},
\end{equation}
\begin{equation}
{\bf j_\epsilon}  = \mathrm{Tr} \big \{ \hat \tau_3 \big( \hat g^R_0 {\boldsymbol \partial_{\bf r}} \hat{g}^R_0 - \hat g^A_0 {\boldsymbol \partial_{\bf r}} \hat{g}^A_0\big )\big \}.
\end{equation}

Since $\tau_{in}^{-1}\ll E_{T}$ the scattering integrals $I$ and $I_{1}$ have a standard form (see Refs.~\cite{LarkinOvchinnikov} and \cite{Aronov}) which 
can be obtained by substituting Eq.~\eqref{gkparam} into the corresponding expression for $\Sigma_{in}$.
They vanish when $f(\epsilon) = \tanh(\epsilon/2T)$ and $f_1 = 0$.

The current density can be expressed in terms of the Keldysh Green's function,
\begin{equation}\label{eq:J(rt)}
{\bf j} ({\bf r, t}) = -\frac{e \tilde \nu_N v_F}{4} \int^\infty_{-\infty} d \epsilon \int \frac{d\Omega_n}{4\pi}\mathrm{Tr}\big \{ \hat \tau_3  \hat g^K(\epsilon,{\bf  r},t, {\bf n})\big \}{\bf n},
\end{equation}
where $\int \frac{d\Omega_n}{4\pi}$ indicates an integration over the direction of the momentum.
Substituting the Keldysh component of Eq.~\eqref{g1 relation} into  Eq.~\eqref{eq:J(rt)} we get an expression for the current density
${\bf j} ={\bf j_d} + {\bf j_{nd}}$, where ${\bf j_d}$ and ${\bf j_{nd}}$ are given by,
\begin{equation}
\label{j_d}
{\bf   j_d} ({\bf r}, t) =  \frac{eD\tilde \nu_N }{4} \int^\infty_{-\infty} d \epsilon  {\bf j_\epsilon} ({\bf r}, t) f(\epsilon,{\bf r}, t),
\end{equation}
\begin{eqnarray}
\begin{split}
\label{j_nd}
{\bf j_{nd}}({\bf r}, t)  =   \frac{e \tilde \nu_N D }{8}\partial_t {\bf A}({\bf r}, t) \int^\infty_{-\infty} d \epsilon \bigg(
\Pi_2(\epsilon,{\bf r}, t) {\bf \nabla_r} f_1(\epsilon,{\bf r}, t)  + \\
 \mathrm{Tr}\bigg \{
2f_1 \hat\tau_3 (\hat g^R_0 \partial_\epsilon \hat g^R_0- \hat g^A_0 \partial_\epsilon \hat g^A_0 ) 
+ \partial_\epsilon f(1 - \hat g^R_0 \hat g^A_0) \\
+ \partial_\epsilon f_1(1 - \hat \tau_3 \hat g^R_0 \hat \tau_3 \hat g^A_0)
+ f \hat \tau_3 (\hat g^R_0 \hat \tau_3 \partial_\epsilon g^R_0 - g^A_0 \hat \tau_3 \partial_\epsilon g^A_0 )\\
+ \partial_\epsilon f \hat \tau_3 (g^R_0 \hat \tau_3 g^R_0 - \hat g^R_0 \hat \tau_3 \hat g^A_0)
+  \partial_\epsilon f_1 \hat \tau_3 (1 - g_0^R g_0^A )
\bigg\}  \bigg).
\\
\end{split}
\end{eqnarray}
The current conservation equation,
\begin{equation}
{\bf \nabla_r} \cdot {\bf j_\epsilon}({\bf r}, t) = 0,
\end{equation}
can also be derived by multiplying Eq.~\eqref{usadel greens} by $\hat \tau_3$ and taking the trace in Nambu space. 

In summary, we have derived a set of equations which describe the kinetics of superconductors in the diffusive regime. The density of states is determined by Usadel's equation \eqref{usadel greens}, the distribution functions are determined by \eqref{diffusive f} and \eqref{diffusive f1}, and the expression for the current is given by \eqref{j_d} and \eqref{j_nd}.

\subsection{Application of the general scheme to the case of SNS Junctions.}

We consider the case where the interaction constant, and consequently the value of the order parameter inside the normal region of the SNS junction is zero.
In this case we can use the following parametrization for the Green's functions,
\begin{equation}
\label{g0 nambu}
\hat g^R_0 = \begin{pmatrix}
G_0^R & F^R_0 \\
-F^{R +}_0 & -G_0^R 
\end{pmatrix}.
\end{equation}
Taking into account that the order parameter inside the normal metal region is zero, we get from Eqs.~\eqref{g0},\eqref{usadel greens} Usadel's equations in the form
\begin{equation}
\label{usadel1}
{\bf \nabla_{\bf r} } \cdot  \bigg(G^R_0 {\bf \nabla_{\bf r}} G_0^R - F_0^R ({\bf \nabla_r} + 2ie{\bf A}) F^{R+}_0 \bigg) = 0,
\end{equation}
\begin{equation}
\label{usadel2}
i\epsilon F^R_0 + \frac{D}{2} \bigg({\bf \nabla_r} - 2ie{\bf A} \bigg) \cdot \bigg(G_0^R ({\bf \nabla_r}- 2ie{\bf A}) F^R_0 - F^R_0 {\bf \nabla_r} G^R_0 \bigg) =0,
\end{equation}
\begin{equation}
\label{normalization gr2}
\big( G_0^R \big) ^2 + F^R_0 F^{R+}_0 = 1 .
\end{equation}
We note that at $T\ll \Delta$ , and  if $L<\sqrt{D\tau_{in}}$  , the distribution function $f_{1}=0$ vanishes everywhere in the sample.
At small voltages $eU \ll E_T$ and in the case of closed boundaries (As it us shown in Fig.~\ref{SNSFIG}) the distribution function $f(\epsilon, t)$ is spatially uniform.

It is convenient to choose a gauge where $\chi=0$, and $\phi = 0$,
\begin{align}
   {\bf p_s}({\bf r},t) = - e {\bf A}({\bf r},t), && {\bf E}({\bf r},t) =  - \partial_t {\bf A}({\bf r}, t),
\end{align}
where ${\bf p_S}$ is the super-fluid momentum and ${\bf E}$ is the electric field. Then Eq.~\eqref{diffusive f} simplifies to
\begin{equation}
\label{diffusive f 2}
\tilde \nu(\epsilon, {\bf r},t) \partial_t  f(\epsilon, t) + \frac{1}{4}e D \tilde \nu_N \partial_\epsilon f(\epsilon, t)   \partial_t {\bf A}  ({\bf r},t) \cdot {\bf j_\epsilon}({\bf r},t)
= I_1 \{f \}.
\end{equation}
Integrating  Eq.~\eqref{diffusive f 2} over the volume of the normal region of the junction we get
\begin{equation}
\label{diffusive f 3}
\begin{split}
\nu(\epsilon,t) \partial_t  f(\epsilon, t) - \frac{ eD \tilde \nu_N S }{4} \partial_\epsilon f(\epsilon,t)U \big({\bf j_\epsilon }(t) \big)_x
= I_1 \{f \},
\end{split}
\end{equation}
where $S$ is the cross sectional area of the junction.
Note that  $j_\epsilon$ must be spatially uniform in this geometry due to the fact that ${\bf j_\epsilon}$ can only depend on the $x$ coordinate and also has a vanishing divergence.

Next we use the diagonal components of Eq.~\eqref{g1 relation} and write $j_\epsilon$ in the following form
\begin{align}
\label{j epsilon}
{\bf j_\epsilon}(t) &= \frac{-v_F}{3D}\mathrm{Tr} \bigg\{ \hat \tau_3 \big( \hat {\bf g}^R_1(\epsilon,{\bf r} ,t) - \hat{\bf g}^A_1(\epsilon,{\bf r} ,t) \big) \bigg \}\\ \nonumber&
= \frac{-2v_F}{D} \mathrm{Tr} \bigg \{ \hat \tau_3  \int \frac{d\Omega_n}{4\pi} \mathrm{Re} \{ \hat g^R (\epsilon, {\bf r},t, {\bf n})\} {\bf n} \bigg \}  \\ \nonumber&
=\frac{2}{m\pi D}  \mathrm{Im} \bigg ( \int d^3{\bf p} \mathrm{Tr} \bigg \{ \hat \tau_3\hat G^R (\epsilon, {\bf r}, t, {\bf p}) \bigg\} {\bf p} \bigg ). 
\end{align}
Differentiating  of both sides of Eq.~\eqref{j epsilon} over $\epsilon$ and integrating over the volume we get,
\begin{equation}
\label{j epsilon2}
\partial_\epsilon {\bf j_\epsilon}(t) =  \frac{2}{\pi DSL} \partial_\epsilon  \mathrm{Im} \bigg (\int d^3{\bf r}   \int d^3{\bf p} \mathrm{Tr} \bigg \{ \hat G^R (\epsilon, {\bf r}, t, {\bf p}) \frac{\bf p}{m}\hat \tau_3\bigg\}  \bigg ).
\end{equation}
Using the fact that $\frac{d \hat H}{d {\bf A}} = -\frac{ie\bf p}{m} \hat \tau_3  + \frac{2ie^2}{m} {\bf A} $, and that ${\bf p} \ll e{\bf A}$ in the quasi-classical approximation, we can write Eq. \eqref{j epsilon2} in the form,
\begin{equation}\label{eq:j}
\partial_\epsilon {\bf j_\epsilon} =  \frac{2i}{e\pi DSL} \partial_\epsilon  \mathrm{Im} \bigg (\int d^3{\bf r}   \int d^3{\bf p} \mathrm{Tr} \bigg \{ \hat G^R (\epsilon, {\bf r}, t, {\bf p}) \frac{d \hat H}{d {\bf A}} \bigg \}\bigg ) .
\end{equation}

To proceed further, we need to derive the following  identity relating derivatives of the Green's functions. 
\begin{equation}\label{eq:identity}
 \int d^3 {\bf r} \int d^3 {\bf p} \mathrm{Tr} \bigg \{ \hat \tau_3 \frac{d\hat G}{d \lambda} \bigg \}  = i\partial_\epsilon  \int d^3 {\bf r} \int d^3 {\bf p} \mathrm{Tr}\bigg \{ \hat G  \frac{d \hat H}{d\lambda}\bigg \}.
\end{equation}
In order to derive this identity, first consider a Hamiltonian and corresponding Green's function with some parametric dependence on $\lambda$,
\begin{equation}
\hat {\mathcal{G}}(\epsilon, \lambda)= \frac{1}{i\epsilon \hat \tau_3 - \hat {\mathcal{H}}(\epsilon,\lambda)}.
\end{equation}
Here $\hat {\mathcal{H}}$ is the Hamiltonian with a particular impurity potential, and $\hat {\mathcal{G}}$ is the exact Green's function of this Hamiltonian.
Calculating the mixed derivatives of the spectral determinant by performing the derivatives $\partial_\epsilon$ and $\partial_\lambda$ in opposite orders, we have the following relations,
\small
\begin{equation}
\partial_\lambda \partial_\epsilon  \int d^3 {\bf r} \int d^3 {\bf p}  \mathrm{Tr} \bigg( \ln \big( \hat {\mathcal{G}}^{-1} \big) \bigg)
= \partial_\epsilon \partial_\lambda   \int d^3 {\bf r} \int d^3 {\bf p} \mathrm{Tr} \bigg( \ln \big( \hat {\mathcal{G}}^{-1} \big) \bigg),
\end{equation}
\normalsize
\begin{equation}
\partial_\lambda  \int d^3 {\bf r} \int d^3 {\bf p}  \mathrm{Tr} \big( \hat {\mathcal{G}}\partial_\epsilon \hat {\mathcal{G}}^{-1} \big)
= \partial_\epsilon  \int d^3 {\bf r} \int d^3 {\bf p} \mathrm{Tr} \bigg( \hat {\mathcal{G}} \partial_\lambda \hat {\mathcal{G}}^{-1}\bigg),
\end{equation}
\begin{equation}\label{spec det}
\partial_\lambda  \int d^3 {\bf r} \int d^3 {\bf p}  \mathrm{Tr}\big( \hat \tau_3 \hat {\mathcal{G}}\big)
= i\partial_\epsilon  \int d^3 {\bf r} \int d^3 {\bf p}  \mathrm{Tr} \big( \hat {\mathcal{G}} \partial_\lambda \hat {\mathcal{H}}\big).
\end{equation}
Next we average Eq.~\eqref{spec det} over impurity configurations. In the case where $\partial_\lambda \hat H$ is independent of the impurity potential, we have equation \eqref{eq:identity}. Using the  Eqs.~\eqref{eq:j} and \eqref{eq:identity}, in the case of $\lambda \equiv {\bf A} $, we have,
\begin{equation}\label{eq2:j}
\partial_\epsilon {\bf j_\epsilon}
 = \frac{2}{e\pi DSL}  \mathrm{Im} \bigg (\int d^3{\bf r}   \int d^3{\bf p} \mathrm{Tr} \bigg \{ \hat \tau_3 \frac{d \hat G}{d {\bf A}} \bigg \}\bigg ) 
=\frac{2}{ e D \tilde \nu_NS } \frac{1}{L} \frac{d \nu}{d {\bf A}}.
\end{equation}

Integrating Eq.~\eqref{eq2:j} with respect to $\epsilon$ and using the fact that ${\bf j_\epsilon}$ is a spatially independent vector which points in the x-direction, we have
\begin{equation}
\label{j epsilon3}
\big( {\bf j_\epsilon} (t) \big)_x =
\frac{- 4}{ S D\tilde \nu_N} \int^\epsilon_0 d \tilde \epsilon \partial_\chi \nu(\epsilon, t) = \frac{ 4}{ S D\tilde \nu_N} \nu(\epsilon) V_\nu(\epsilon),
\end{equation}
where $\chi(t) =-2e\int^{L/2}_{-L/2}  dx {\bf A}_x (x,t)$ is the phase difference across the junction.
Substituting Eq.~\eqref{j epsilon3} into \eqref{diffusive f 3} we reproduce Eq.~\eqref{EQ:N_DOT} in the main text.
\begin{equation}
\label{kinetic_equation}
\partial_t     f(\epsilon, t)  + 2 eU \partial_\epsilon f (\epsilon, t) V_\nu (\epsilon) = I_1 \{ f\}.
\end{equation}

\subsubsection{Expression for the current}

Let us consider the equation for  the diagonal current ${\bf j_d}$ at small voltages when the distribution function $f$ is specially uniform.
\begin{equation}
\label{current j epsilon}
J_d =\frac{eD\tilde \nu_N S }{4 L} \int^\infty_{0} d \epsilon f(\epsilon)  \int^{L/2}_{-L/2} dx   \big( {\bf j_\epsilon}(t) \big)_x .
\end{equation}
Substituting Eq.~\eqref{j epsilon3} into Eq.~\eqref{current j epsilon}  we get,
\begin{align}\label{current_f}
J_d & = e \int^\infty_{0} d\epsilon \nu(\epsilon)  f(\epsilon)  V_\nu(\epsilon) & \nonumber \\
& \equiv J_c(0) Y(\chi,0) - e \int^\infty_0 d\epsilon \nu(\epsilon) V_\nu(\epsilon)(1 - f(\epsilon)).&
\end{align}
Using the relationship, $n(\epsilon) = \frac{1}{2}\big(1- f(\epsilon) \big), \,\ \epsilon>0 $, we see that Eq.~\eqref{current_f} is equivalent to the expression for the diagonal current used in the main text(see Eq.~\eqref{EQ:CURRENTEUL}).

Let us now turn to the non-diagonal contribution to the current, $j_{nd}$.  To linear order in ${\bf E}$ an estimate for $j_{nd}$ can be obtained by substituting the equilibrium distributions into Eq.~\eqref{j_nd},
\begin{equation}
\label{nd current}
\begin{split}
{\bf j}_{nd} ({\bf r}, t)  =  \frac{e \tilde \nu_N D }{2} {\bf E}({\bf r},t) ({\bf r}, t) \int^\infty_{0} d \epsilon \bigg[
\tanh(\epsilon/2T)  \partial_\epsilon \bigg( (G^R_0)^2 + F^R_0 F^{R+}_0 \\- (G^A_0)^2 - F^A_0 F^{A+}_0 \bigg)
+ \frac{2}{T}\frac{(G^R_0)^2 + |G^R_0|^2 }{\cosh^2(\epsilon/2T) }
\bigg].
\end{split}
\end{equation}
The dominant contribution to the integral comes from the region when $\epsilon \sim T$. In the case when $T \gg E_T$, the Green's functions are equal to the normal metal Green's functions in the relevant energy intervals. In this case
\begin{equation}
\label{nd current 2}
\begin{split}
{\bf j}_{nd} ({\bf r}, t)  \sim  G_N {\bf E}({\bf r},t).
\end{split}
\end{equation}
Thus we have shown that the non-diagonal current is of the same order as the dissipative current in the normal state.

\end{document}